\documentclass{article}

\PassOptionsToPackage{numbers,compress}{natbib}
\usepackage[preprint]{neurips_2026}   

\usepackage[utf8]{inputenc} %
\usepackage[T1]{fontenc}    %
\usepackage{hyperref}       %
\usepackage{url}
\usepackage{tabularx}%
\usepackage{booktabs}       %
\usepackage{amsfonts}       %
\usepackage{nicefrac}       %
\usepackage{microtype}      %
\usepackage{xcolor}         %
\usepackage{graphicx}
\usepackage[dvipsnames]{xcolor}
\usepackage{amsmath}
\usepackage{amssymb}
\usepackage{enumitem}

\newcommand{\R}{\mathbb{R}}
\newcommand{\geom}{\text{geom}}
\newcommand{\meme}{\text{MEME}}

\newcommand{\f}[1]{f/\texttt{#1}}

\title{Interpreting Latent Protein Language Model Features with Geometric Annotations}

\author{
  Siddharth Setlur$^{\ast}$ \\
  School of Mathematics \\
  University of Edinburgh, UK \\
  \And
  Djordje Mihajlovic\thanks{Joint first author.} \\
  School of Mathematics \\
  University of Edinburgh, UK \\
  \AND
  Darrick Lee \\
  School of Mathematics \\
  University of Edinburgh, UK \\
}

\begin{document}

\maketitle

\begin{abstract}

Protein language models (pLMs) encode information about protein sequences which enable downstream tasks such as structure prediction, but their internal representations are not well understood. Sparse autoencoders (SAEs) provide a promising tool to disentangle latent pLM representations into interpretable features, but existing annotation pipelines largely rely on protein-level annotations derived from database labels and LLM annotations of top activating sequences. Such annotations can overlook the localized residue-level and geometric patterns encoded by sparse features.
We introduce an automated and scalable method for interpreting SAE features in ESM-2 by using geometrically inspired features of the protein $\text{C}_{\alpha}$ backbone. Across ESM-2 8M layers, an FDR-controlled discovery analysis shows that local geometry is significantly associated with many SAE features, with varying levels of predictive strength,
expanding coverage beyond database and sequence-based methods. In particular, geometry can distinguish SAE features sharing the same database annotation, revealing substructure within known biological labels. A significant portion of SAE features activate on unannotated metagenomic protein sequences  enabling us to use our SAE annotations to better understand these sequences. In addition, ablation experiments at the level of contact prediction show that removing found geometric features shifts ESM-2’s predicted contact maps in the direction of the descriptor.
This provides a robust method of annotating proteins activated within SAE neurons at a residue level, providing a bridge between mechanistic interpretability and structural biology.

\end{abstract}

\section{Introduction}\label{sec:intro}

Protein language models (pLMs) adapt transformer architectures \cite{vaswani_attention_2017} to encode the \textit{language of proteins}. They are trained on large datasets of protein sequences\footnote{Protein sequences use the 22 amino acid positions as tokens; for example, a fragment of the sequence for Insulin is MALWMRLLPLLA\ldots \cite{ahmad_uniprot_2025}. A single character (amino acid) in a protein sequence is called a \emph{residue} in the protein literature.} and learn representations of these sequences to perform a variety of crucial tasks including predicting protein structure \cite{lin_evolutionary-scale_2023} and function \cite{chen_evaluating_2025}. 
The success of these structure prediction models such as ESMFold \cite{lin_evolutionary-scale_2023} is particularly impressive because training is primarily on unaligned sequences, in contrast to models like AlphaFold2 \cite{jumper_highly_2021} which require multiple sequence alignments (MSA) as a core part of their training corpus.
Because MSA data are limited due to  relative to the vast amount of available protein sequence data \cite{ahmad_uniprot_2025}, pLM-based structure prediction methods are especially important for unannotated metagenomic protein sequences\cite{lin_evolutionary-scale_2023}, such as uncharacterized microbial enzymes and viral sequences \cite{flamholz_large_2024, thurimella_protein_2025}.

Despite their success in structure prediction and protein annotation~\cite{rives_biological_2021}, it remains unclear what features pLMs learn from unlabelled sequence data, how these features are represented internally, and how they are used to make predictions. Mechanistic interpretability offers a framework to address this gap by identifying human-interpretable features in model representations.
Such interpretations can help improve protein annotation in databases, characterize unannotated metagenomic proteins \cite{baltoumas_nmpfamsdb_2024, pavlopoulos_unraveling_2023}, and provide deeper insight into the principles governing protein folding. 
Interpretability may also support applications such as disease variant prediction and protein design, where understanding model predictions is often crucial for assessing reliability \cite{brandes_genome-wide_2023,ismail_concept_2024}.

\textbf{Sparse Autoencoders for Monosemantic Features.}
A key difficulty in understanding internal representations of neural networks is that individual neurons are \emph{polysemantic}~\cite{olah_zoom_2020}: they may encode a mixture of multiple concepts simultaneously~\cite{elhage_toy_2022}. 
Sparse autoencoders (SAEs) address this by decomposing the internal representations of neural networks into a larger set of sparse latent features, called \emph{SAE features}. 
The hope is SAE features are \emph{monosemantic}, and can be \emph{annotated} with a human interpretable concept by identifying commonalities among the input tokens which produce strong feature activations.
However, due to the size of modern LLMs and pLMs, annotating these features at scale is a challenge. 
For LLMs the annotation is often automated by prompting another language model to infer a human-interpretable description from the top activating tokens~\cite{paulo_automatically_2025, steven_language_2023}.

\textbf{Database and LLM based Annotations for pLM SAE features.}
Recently, \cite{simon_interplm_2025, adams_mechanistic_2025, silberg_towards_2025} applied these techniques to identify and annotate monosemantic features of ESM-2 embeddings including biological function domains, sequence motifs, and binding sites.
These pipelines query databases such as InterPro \cite{blum_interpro_2025} which aggregates resources such as UniProtKB~\cite{ahmad_uniprot_2025} (functional), Pfam (domain)~\cite{paysan-lafosse_pfam_2025}, and CATH/Gene3D~\cite{sillitoe_cath_2021} (hierarchical structure) providing \emph{database annotations} for the subset of SAE features whose top-activating proteins share known database labels.
While these annotations provide valuable biological context at the protein level, SAE features often encode more localized residue-level patterns.
For the remaining SAE features, LLM-generated descriptions of top-activating proteins provide an alternate method, but they depend on existing literature on the corresponding proteins and may not fully exploit structural information.

\textbf{Contributions.}
Motivated by the need for annotations that are both residue-localized and derived from protein sequence and structure, we introduce an automated and scalable pipeline for annotating pLM SAE features at \emph{residue-level} resolution. Our main contribution is a geometrically informed annotation method that tests whether SAE activations are predictable from local $C_\alpha$ backbone geometry, supplemented by amino-acid covariates that capture basic physicochemical priors. While the individual components of the framework are established, our contribution lies in the residue-level, structure-derived annotations produced by their integration and the analyses these annotations enable. The full pipeline also includes residue-level database annotations inspired by~\cite{simon_interplm_2025}, MEME-based sequence motif annotations~\cite{bailey_meme_2015} and positional annotations, providing complementary methods for assigning interpretable descriptions to SAE features. Empirically, we show that our pipeline:
\begin{enumerate} [itemsep=-2pt, topsep=3pt]
    \item annotates SAE features missed by InterPro database annotations alone;
    \item separates SAE features sharing the same InterPro label into finer geometric motifs;
    \item transfers geometric annotations from SwissProt to some metagenomic proteins;
    \item links geometry-annotated features to contact-map changes under ablation.
\end{enumerate}

We also release \href{https://geopedia.studio}{Geopedia}\footnote{https://geopedia.studio}, an interactive interface for examining the full set of SAE annotation results across features and layers. Code to reproduce our results can be found at our \href{https://github.com/siddharthsetlur/ProteinLens}{Github repo}\footnote{https://github.com/siddharthsetlur/ProteinLens}.

\textbf{Related Work.}
Prior work on interpretability of pLMs suggests that the internal representations of these models encode biologically meaningful information. Specifically, several papers explicitly find correlations between residue contacts, binding sites, and evolutionary couplings at the level of attention and representation of pLMs~\cite{rao_transformer_2020, vig_bertology_2020, valeriani_geometry_2023}. 
Beyond this, SAEs have been used to navigate and annotate the internal representations more directly, often finding that protein family, domain, and functional annotation are highly correlated with SAE feature~\cite{garcia_interpreting_2025, parsan_towards_2025, liu_protsae_2026, gujral_sparse_2025}.

\section{Background}

\textbf{Protein Language Models.} 
A protein language model maps a protein sequence to a sequence of internal residue representations. In this work, we focus on ESM-2 \cite{lin_evolutionary-scale_2023}, an encoder-only transformer trained on the UniRef50 databank of protein sequences~\cite{ahmad_uniprot_2025}. Given an input protein sequence, ESM-2 embeds each residue as a token and updates these token representations thorugh a stack of bidirectional encoders~\cite{devlin_bert_2019}. In particular, the hidden state at each residue can depend on sequence context from both sides, and each layer produces a sequence of residue-level embeddings.

ESM-2 is trained as a masked language model: a random subset of residues, approximately $15\%$, is hidden, and the model is trained to predict the masked amino acids from the remaining unmasked context. In our experiments, we use ESM-2 8M which has six transformer layers and $320$-dimensional residue embeddings.
While these models are trained exclusively on sequence data, ESM-2 representations encode information about protein structure and function. In particular, such masked language models are known to recover residue-residue contact maps of proteins~\cite{rao_transformer_2020, vig_bertology_2020}. This motivates our central question: whether latent features in ESM-2 can be annotated at residue-level resolution, and whether some of these features correspond to local geometric structure.

\textbf{Sparse Autoencoders.} 
Sparse autoencoders are used to decompose the dense latent representations of models into sparse, more interpretable features.  Given an activation vector $x^{(l)} \in \R^m$ from layer $l$, an SAE learns an overcomplete feature representation $f(x^{(l)}) \in \R^F$, where $F \gg m$, using an encoder $E: \R^{F \times m}$ and a decoder $D: \R^{m \times F}$, such that
\begin{align*}
    \hat{x}^{(l)} = b_D + E f(x^{(l)})  \quad \text{where} \quad f(x^{(l)}) = \text{ReLU}(Ex^{(l)} + b_E),
\end{align*}
and $b_E \in \R^F$, $b_D \in \R^m$ are the encoder and decoder bias terms.
The model is trained to minimize 
\begin{align*}
    \mathcal{L}_{\text{SAE}}(x^{(l)}, \hat{x}^{(l)}) = \| x^{(l)}-\hat{x}^{(l)}\|_2^2 + \lambda\|f(x^l)\|_1
\end{align*}
where the first term ensures faithful reconstruction of the activation vector, while the second term enforces sparsity so that only a small number of the components of $f(x^{(l)})$ are nonzero.
These components (the coordinates of $\R^F$) are called \emph{SAE} features, and our aim is to annotate them with monosemantic labels. 
While other variants of SAEs have been introduced to address issues such as feature absorption~\cite{bussmann_learning_2025, gao_scaling_2024}, we use the basic RELU SAE architecture~\cite{bricken_towards_2023} described above.
\label{sec:sae-background}
\begin{figure}[t]
    \centering
    \includegraphics[width=0.9\linewidth]{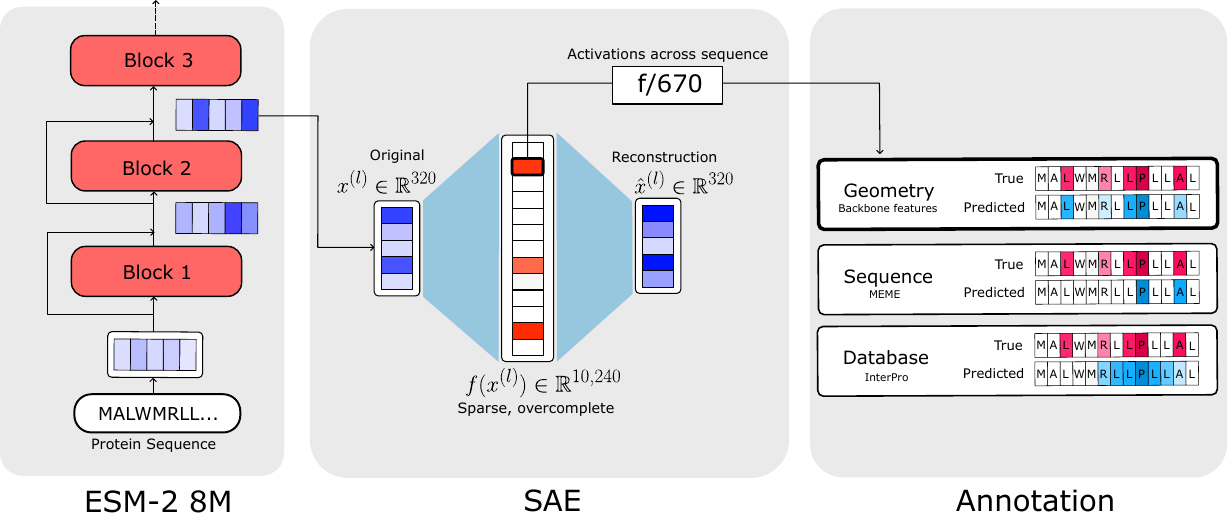}
    \caption{Sparse autoencoder trained on the residual stream of ESM2-8M.
    The SAE encoder maps the 320-dimensional residual into 10{,}240 sparse
    features ($32\times$ overcomplete), and the decoder reconstructs the
    original activation.}
    \label{fig:sae-arch}
\end{figure}

\textbf{Training and Inference Data.} 
For each of the six residue layers of ESM-2 8M, we train a separate ReLU SAE on residue-level activations extracted from a corpus of proteins from the SwissProt dataset.
We consider SAEs with a $10,240$-dimensional features ($32$ times the $320$-dimensional ESM-2 embedding).
This produces a collection of sparse features independently at every layer, each of which can activate at individual residue positions. Details of the SAE training data, hyperparameter sweep, and model selection are given in \autoref{sec:appendix-sae-training}. 

We then run SAE inference on $50,000$  protein sequences to determine which sequences a given SAE feature fires on. To ensure that the sample is diverse,  we cluster all SwissProt sequences with mmseqs easy-cluster \cite{steinegger_mmseqs2_2017} at $30\%$ sequence identity and $80\% $ bidirectional coverage, then sample at most one accession per cluster representative. This ensures that large near-identical families contribute at most one protein each, preventing them from dominating the subset.

\section{Residue-Level Annotation for pLM SAE Features}
We integrate established annotation and prediction methods into a scalable pipeline for annotating SAE features of pLMs at the \emph{residue-level}.
Our main contribution is a geometrically informed annotation method, which tests whether SAE activations are predictable from local features of the protein $C_\alpha$ backbone. We consider three further complementary classes of annotations: database annotation, sequence motifs, and sequence positions. The database method is a residue-level analogue of~\cite{silberg_towards_2025} while the sequence-motif method uses uses tools methods similar to \cite{hou_motifae_2025}. Each annotation method produces either a continuous score or binary predicate at the residue level, which we evaluate against SAE activations using a common significance-testing procedure. 

\textbf{General Annotation Pipeline.} The general pipeline for annotations for a fixed layer $l$ and a fixed SAE feature $\f{r}$ is built using the following structure, with minor differences between continuous (c) and binary (b) annotation methods.
For a protein sequence $s = (s_1, \ldots, s_L)$, let $x_i \in \R^m$ denote the ESM activation vector at layer $l$ and residue position $i$, and $a_{r,i}(s) = f_r(x_i) \in \R$ denote the corresponding activation of SAE feature $\f{r}$. We provide full details of these steps in \autoref{sec:appendix-annotation-pipeline}.
\begin{enumerate}
    \item \textbf{Select proteins.} For each SAE feature $\f{r}$, we choose a set $S_r$ of proteins, whose size may vary by feature. We select proteins from the top activating proteins (c), or by sampling across normalized activation levels (b). For each $s \in S_r$, we record $a_{r,i}(s)$ for all residues. 
    \item \textbf{Define active residues.} We threshold normalized SAE feature activations by $\gamma \in [0,1]$ to obtain binary residue labels $y^\gamma_{r,i}(s) \in \{0,1\}$ indicating where $\f{r}$ fires. The activation threshold $\gamma$ is either fixed in advance (c) or computed over a grid of values (b). 
    \item \textbf{Compute annotation predictions.} Each annotation method $a$ provides either a continuous score $\hat{y}^a_{r,i}(s) \in \R$ or a binary predicate $\hat{y}^a_{r,i}(s) \in \{0,1\}$ at each residue for every $s \in S_r$, computed using its corresponding source of information: local backbone geometry, database labels, sequence motifs, or sequence position.
    \item \textbf{Evaluate and test significance.} We compare the annotation predictions $\hat{y}^a_{r,i}(s)$ against the binary SAE activations $y^\gamma_{r,i}(s)$ either using PR-AUC on the single threshold (c) or F1 scores, then optimizing $\gamma$ (b). We then assess significance against a residue-shuffled null distribution with multiple-testing correction. 
\end{enumerate}

\subsection{Geometrically-Informed Annotation}
The central method proposed in this article is a geometrically-informed annotation method which tests whether an SAE feature fires on the premise of shared $\text{C}_{\alpha}$ backbone geometry~\cite{ramachandran_stereochemistry_1963, oldfield_analysis_1994, durairaj_geometricus_2020}, along with weak sequence context to dictate properties such as flexibility and distant contact restrictions (e.g. cysteine bonds~\cite{wiedemann_cysteines_2020}). This is motivated by evidence that pLMs often encode structural information such as residue-residue contacts~\cite{rao_transformer_2020, vig_bertology_2020}.

\textbf{Local Geometric Features.} For each protein $s \in S_r$, we retrieve its  AlphaFold DB structure and extract the $\text{C}_{\alpha}$ backbone.
Around each residue $i$ we form a local window
$W_{h}(i) = \{i-h, ..., i+h \}$ with $h=10$, corresponding to 21-residue fragments respectively.
For each window, we compute a geometric feature vector, $\phi_{i}(s) \in \mathbb{R}^{44}$ consisting of local curvature, torsion, planarity, contact statistics, and coarse amino-acid composition covariates. These amino-acid covariates are included in the feature vector as weak physicochemical priors: they capture broad chemical constraints on the protein backbone that result in particular geometries. See \autoref{sec:appendix-geometric-feats} for the full list of features. 

\textbf{Annotation Prediction.} For each SAE feature $\f{r}$, we train a gradient boosted classifier (GBM) to predict the active residue-level activations $y_{r,i}(s)$ using geometric feature vector $\phi_{i}(s)$. This classifier outputs a continuous geometric annotation score $\hat{y}^{\geom}_{r,i}(s)$, which estimates whether feature $\f{r}$ fires at residue $i$ based on $\phi_i(s)$. We evaluate this score using PR-AUC, following the continuous annotation procedure described above. 

This approach complements the structural analysis of~\cite{silberg_towards_2025}, which compare 100-residue structural neighbourhoods between top-activating proteins to identify shared structural similarity using RMSD.
In contrast, our method is residue-level: rather than comparing structural alignment, we test whether the residues on which an SAE feature activates are predictable from local geometric descriptors. Note that the training data is accessed via AlphaFold across the SwissProt dataset; here it is important to note that the median pLDDT is $91.35$~\cite{schaeffer_ecod_2026}.

\subsection{Complementary Annotation Baselines}
We also evaluate three complementary annotation baselines: database annotations, sequence motifs, and sequence position. Each baseline uses the same feature-wise prediction framework described above, but restricts the predictor to a different source of information. 

\textbf{Database Annotations.}
We extend the approach detailed in \cite{silberg_towards_2025} to use the InterPro\cite{blum_interpro_2025} annotations as residue-level biological baselines. For each candidate database label $\text{b}$, we define a binary predicate $\hat{y}^{\text{b}}_{h,i}(s) \in \{0,1\}$ which indicates whether residue $i$ of protein $s \in S_r$ lies inside a residue interval associated with $\text{b}$. We then compare this against the SAE active-residue labels $y_{r,i}(s)$ using F1 score, selecting the candidate label with the higher score for each SAE feature $\f{r}$.

\textbf{Sequence Motifs.}
To test whether SAE activations are explained by local sequence patterns, we apply Multiple EM for Motif Elicitation (MEME) \cite{bailey_meme_2015}, a classical tool in bioinformatics, to sequence windows around high-activation residues. 
The resulting motifs are represented as position weight matrices (PWM). We convert their entries to log-odds against a feature-specific amino-acid background and define $\hat{y}_{r,i}^{\meme}(s) \in \R$ as the summed log-odds score at that residue, using the motif with the highest PR-AUC for feature $\f{r}$.
This baseline separates geometric explanations from purely sequence motifs. See \autoref{sec:appendix-meme} for further details.

\textbf{Sequence Position.} 
Finally, we include position predicates to detect features that fire at systematic locations within proteins; these predicates are listed in \autoref{sec:appendix-position-predicates}. Each candidate position annotation $\text{p}$ defines a binary predicate $\hat{y}^{\text{p}}_{r,i}(s) \in \{0,1\}$, evaluated by F1 score. This baseline controls for positional effects that are independent of amino-acid identity or protein structure.

\subsection{Evaluation and Significance Testing}
\label{sec:background-eval}

The main question we seek to answer is: given that a feature fires strongly on a protein, is the residue-level activation pattern predictable from local backbone geometry.
For each SAE feature $\f{r}$ and each annotation method, we compare annotation predictions $\hat{y}_{r,i}(s)$ to binary active-residue labels $y^\gamma_{r,i}(s)$: continuous methods use a fixed feature-specific threshold $\gamma$ and PR-AUC, while binary methods sweep $\gamma$ over a grid and use the best F1 score. We test significance with a residue-shuffled null model within selected proteins in $S_r$, shuffling active labels within each protein to preserve the number of active residues while destroying alignment with the annotation signal \cite{phipson_permutation_2010}. For each feature-method pair, we generate 100 permutations and recompute the relevant score, yielding a permutation $p$-value. 
Because we test 10,240 SAE features for each annotation method, we apply the Benjamini--Hochberg (BH) correction separately within each annotation method and layer \cite{benjamini_controlling_1995}. This correction adjusts the permutation $p$-values such that among the features called significant for a given annotation method, the expected fraction of false discoveries is controlled. We report these corrected $p$-values, called $q$-values, and call a feature $\f{r}$ \emph{significantly annotated} by a method when $q < 0.05$. An annotation method is called \emph{primary} if it is the only annotation method with $q < 0.05$. This procedure controls the expected false-discovery rate (FDR) among selected features, but does not control the probability of making any false discovery across the family of tests. For the geometric annotation, our primary permutation test fixes the trained GBM scores and shuffles residue labels within each protein; this is reported in \autoref{sec:results}. 
Retraining GBMs for each permutation would provide a stricter null, which accounts for the model-fitting step itself, but it is computationally prohibitive across all permutations, layers and features. Thus, we perform this retraining-based test on layer $4$ as a representative robustness check, which reduces $q$-significant features by $\approx 0.3\%$, suggesting that the fixed-score permutation test is not driving the observed significance.

\section{Results}\label{sec:results}

We evaluate whether our residue-level annotation pipeline can assign interpretable descriptions to SAE features in ESM-2 8M, and whether geometry provides information beyond existing database and sequence based annotation. We organize the results around four questions. 

\begin{enumerate}[itemsep=-1pt, topsep=2pt]
    \item How much of the SAE feature space is covered by each annotation method?
    \item Can local geometry reveal finer substructure within database annotations?
    \item Can geometric annotations transfer to metagenomic proteins without database annotations?
    \item Do SAE feature ablations induce consistent geometric changes in contact predictions?
\end{enumerate}

To support qualitative inspection of these results, we provide an interactive visualization tool for exploring top activating proteins across all SAE features and layers: \href{https://geopedia.studio}{GeoPedia}, allowing readers to inspect both the case studies below and the broader collection of annotated features. This interface is intended to make the results transparent to biologists and domain experts, and to support further biological interpretation of motifs beyond the selected examples. Our code is provided in \href{https://github.com/siddharthsetlur/ProteinLens}{Github Repo}. Our visualizer additionally binarizes the continuous annotations, see \autoref{sec:appendix-binarization}.

\textbf{1. FDR-Controlled Annotation Discovery Across SAE Features.}
A combination of our residue-level annotation methods covers many SAE features across ESM-2 8M layers, with local geometry providing the broadest single-method coverage.
We say an SAE feature is \emph{annotated} when its permutation-test $q$-value is below 0.05, as described in \autoref{sec:background-eval}; total coverage is the union over methods. As a baseline comparison, we compute protein-level database annotations using a procedure similar to \cite{silberg_towards_2025}, described in \autoref{sec:appendix-plbaseline}, and recover comparable statistical significance to their reported values. \autoref{tab:annotated_percentage} summarizes these results.

\begin{table}[hbt!]
    \centering
    \resizebox{\textwidth}{!}{%
    \begin{tabular}{c|c|c|c|c|c|c}
         & \% Total annotated & \% InterPro Prot. & \% InterPro Res. &\% Seq Pos & \% Seq Motif & \% Geometric \\
         Layer 2 & 86.53 & 56.11 & 42.03 & 62.22 & 69.33 & 84.38 \\
         Layer 4 & 93.50 & 56.75 & 61.46 & 79.19 & 77.56 & 92.79 \\
         Layer 6 & 94.57 & 56.68 & 67.70 & 71.35 & 77.20 &  94.49 \\
         
    \end{tabular}
    }
    \caption{ FDR-controlled discovery coverage across annotation methods.
    We say that a feature is annotated by a given annotation method, if its $q$ value if below $0.05$. We say that a feature is annotated if any of the annotation methods have a $q$ value if below $0.05$.}
    \label{tab:annotated_percentage}
\end{table}

    We note that $q$-significance measures departure from the permutation null, not the predictive strength of an annotation. Empirically, SAE features activate on approximately $1\%$ of residues in a given protein, so random classifiers achieve low PR-AUC values, around $0.01$ with median null of $0.007$. Conversely, coarser methods such as protein-level InterPro annotations predict only whether an SAE feature activates on a protein, leading to a higher median null F1 score of approximately $0.5$, making rejection of the null much more difficult.
    For this reason, we focus primarly on geometric annotations whose PR-AUC values are above 0.3. \autoref{tab:PR-AUC} considers the distribution of PR-AUC values of geometrically annotated features.

    \begin{table}[hbt!]
        \centering
        \begin{tabular}{c|c|c|c}
            & PR-AUC 0.0-0.3 & PR-AUC 0.3-0.6 &  PR-AUC $>0.6$ \\
             Layer 2 & 95.15 & 3.27 & 1.58 \\
             Layer 4 & 81.72 & 12.24 & 6.05 \\
             Layer 6 & 91.86 & 6.51 & 1.63 \\
             
        \end{tabular}
        \caption{PR-AUC values of geometrically annotated features on a set of 10k held out proteins across tested layers within ESM-2 8M. The 10k held out set is similarly sampled from the SwissProt database, explicitly disjoint from the 50k training samples.}
        \label{tab:PR-AUC}
    \end{table}

\begin{figure}[hbt!]
    \centering
    \includegraphics[width=0.8\linewidth]{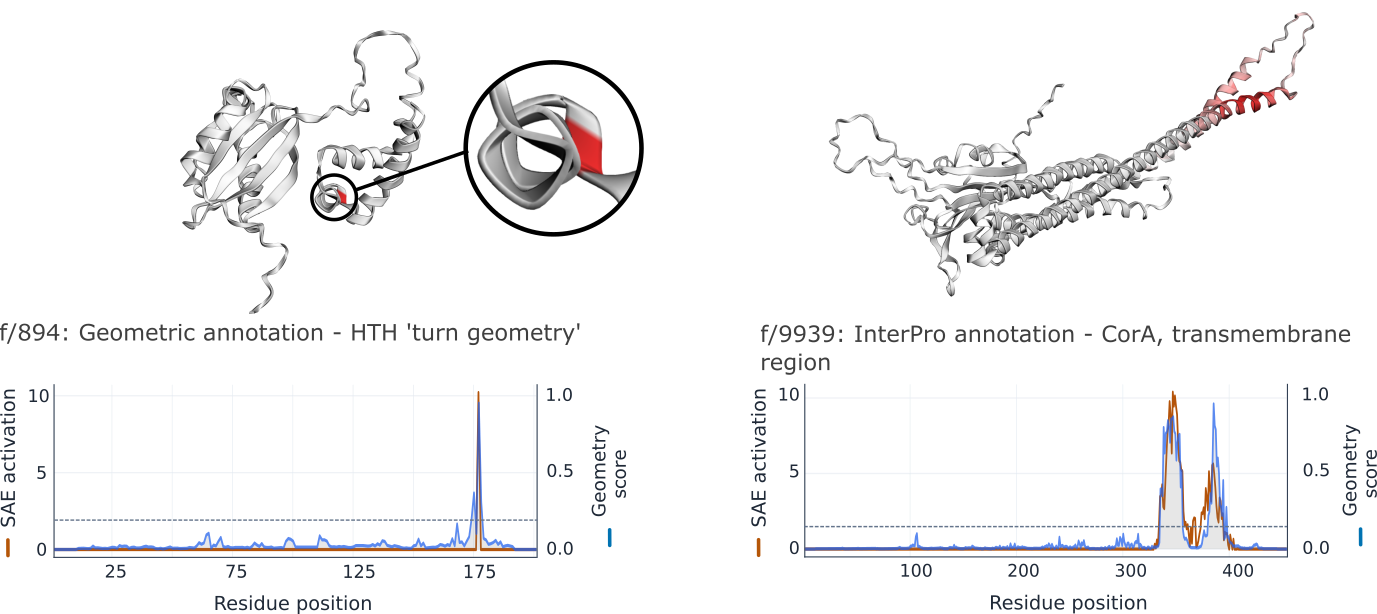}
    \caption{\textbf{Examples of different annotation methods across features.} Comparing the annotations returned against all methods and respective statistical null tests (as in~\autoref{tab:annotated_percentage}) we present SAE features whose annotation is considered geometry-primary (left) and InterPro residue primary (right). It is clear from the select examples and activations at their top activating proteins (plots accessed via~\cite{rego_3dmoljs_2015}) that the dominant annotation type can vary. We place specific emphasis on the feature f/894 whose annotation is only accurately available within the geometric pipeline developed in this work. In the case of the geometry-primary feature, the high GBM activation prediction accuracy \emph{automatically} helps demonstrate that the underlying notion encoded at this feature is likely structural in nature.}
    \label{fig:examples}
\end{figure}

    Across the SAE, some features are primarily explained by a single annotation method.
    To illustrate these annotation regimes qualitatively, \autoref{fig:examples} shows representative SAE features with annotations considered geometry-primary or InterPro residue-primary. These examples show that InterPro annotations often correspond to broad activation over known biological regions, whereas geometric annotations capture more localized structural patterns. At the same time, many features receive overlapping annotations, motivating a closer analysis of whether features with the same InterPro label encode distinct geometric patterns. \autoref{sec:appendix-geom-feat-importance} further demonstrates that high PR-AUC geometric annotations are dominated by geometric $\text{C}_{\alpha}$ backbone features as opposed to amino acid covariates.

    To demonstrate the effect is not localized to ESM-2 8M, we further evaluate our results by running the pipeline across ESM-2 35M, layer 6. The results are highlighted in~\autoref{sec:appendix-35M} where we obtain similar results to the 8M case. Furthermore, we validate our results using experimentally determined protein structures for a small subset of protein-feature pairs in~\autoref{sec:appendix-real_protein}.

\textbf{2. Residue-level geometry reveals substructure within known biological motifs.}
While many features may have overlapping annotations, the \emph{quality} of these annotations varies. Here, we test whether geometric annotations can uncover finer details encoded within SAE featues beyond the coarse InterPro annotations. SAE features often localize to specific residues within known structural motifs, suggesting that sparse features decompose broad domain/family level annotations into finer geometric submotifs. \autoref{tab:case_geom} reports the percentage of features that share InterPro annotations which are distinguishable using the geometric annotation. We compare the GBM feature-importance vectors learned for each feature, and call two featuers \emph{geometrically distinguishable} when the cosine similarity between these vectors is below $0.5$.

\begin{table}[hbt!]
    \centering
    \begin{tabular}{c|c|c}
         & No. InterPro Res families & \% Geom Distinguishable \\
         Layer 2 & 120 & 59.17 \\
         Layer 4 & 578 & 77.51 \\
         Layer 6 & 45 & 68.89 \\
         
    \end{tabular}
    \caption{Geometric annotations further split InterPro residue annotations, revealing more granular features encoded across a subset of SAE features between layers.}
    \label{tab:case_geom}
\end{table}

\begin{figure}[hbt!]
    \centering
    \includegraphics[width=0.8\linewidth]{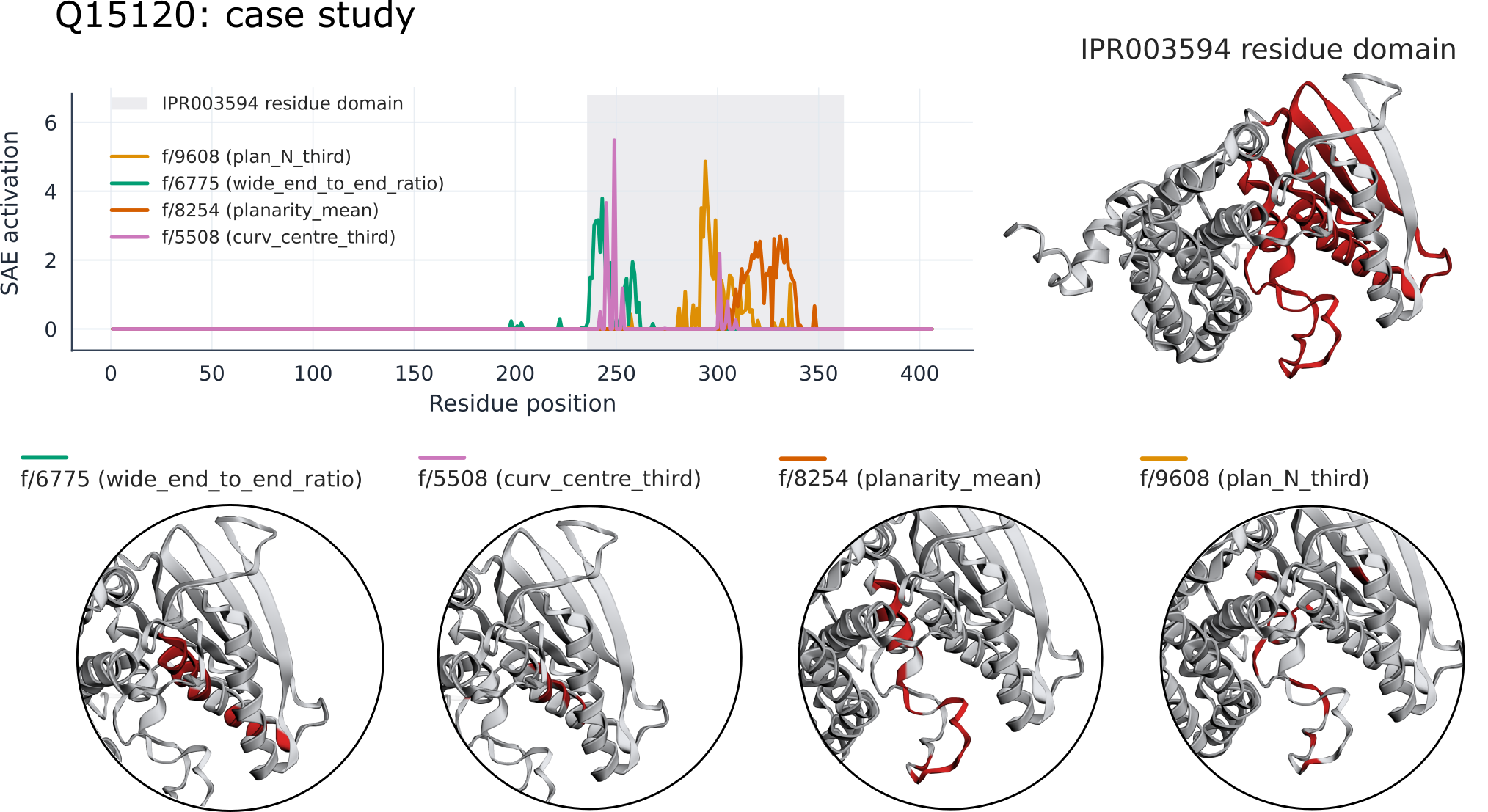}
    \caption{\textbf{Annotation exists beyond coarse InterPro labels.} Here, we compare a set of SAE features whose biological annotation via residue-level InterPro label is consistently `Histidine kinase/HSP90-like ATPase domain'. Activation can be seen to vary substantially across the domain, suggesting finer features beyond the coarse biological annotation, which the geometric annotation method is able to detect. On this specific example, activation occurs across four `sub-domains' within the larger IPR003594 annotation which remain distinct in both geometry space and sequence localization (median cosine 0.25; 80.6\% zero overlap on shared proteins). Note, all GBM geometric feature vector predictions for these activations achieved PR-AUC $>0.65$.}
    \label{fig:shared_bio}
\end{figure}

~\autoref{fig:shared_bio} illustrates this separation for an example at layer 4 for four SAE features $\f{6775}$, $\f{5508}$, $\f{8254}$, and $\f{9608}$, evaluated on the protein Q15120. All four features receive the same high confidence InterPro residue annotation  `Histidine kinase/HSP90-like ATPase domain,' but their activations occupy different subregions of the domain rather than the full annotated interval. 
Qualitative analysis of the activation highlighted plots in~\autoref{fig:shared_bio} demonstrates four distinct geometric motifs within the larger structure.
From left to right in ~\autoref{fig:shared_bio}: $\f{6775}$ activates highly across an beginning and end thirds of $\alpha$-helix rich region at one end of the broader structure, whereas $\f{5508}$ then activates on the center third. The feature $\f{9608}$, can then be seen to activate highly across the connective portions between the two $\alpha$-helix rich regions across the domain and finally $\f{8254}$ captures the subsequent $\alpha$-helix rich region toward the end of the structure. The geometric based annotations are able to annotate each of these subdomains via their GBM predicted feature vector.

Prior work performed a similar refinement manually, using an LLM-assisted literature search to identify more granular subdomains for $6$ of $8$ SAE features~\cite{silberg_towards_2025}. Our geometric method provides a robust and scalable alternative: rather than refining protein-level annotations case by case, we systematically test whether shared residue-level annotations split into local geometric patterns.

\textbf{3. Geometric annotation captures transferable annotation to metagenomic proteins.}
Metagenomic proteins provide a natural test of whether geometric SAE annotations transfer beyond well-annotated SwissProt proteins. Prior work found that SAE features can identify structurally similar regions in the NMPFams  sequences \cite{baltoumas_nmpfamsdb_2024} by matching activated metagenomic regions to characterized SwissProt examples using $\text{RMSD}_{100}$ distances \cite{silberg_towards_2025}. In contrast, we test transfer at the residue level: for each geometrically annotated SAE feature, we evaluate the SwissProt-trained geometric GBM directly directly on metagenomic proteins where that feature activates, and ask whether the same local geometric predictors explain activation in this new setting.

We survey $50,000$ NMPFams metagenomic families and identify SAE features that activate on these unseen sequences. 
For a geometrically annotated feature, high PR-AUC on the activating metagenomic proteins indicates that the same local geometric pattern predicts SAE activation outside the SwissProt annotation set. 
This provides a transfer criterion that does not require existing InterPro annotations, which are often unavailable for metagenomic proteins. 
\autoref{tab:PR-AUC-metagenomic} reports the fraction of features with NMPFams activations, the fraction which are geometrically $q$-significant, and the portion of features with PR-AUC values above $0.5$ for GBMs \emph{evaluated on the metagenomic sequences}. 
Furthermore, column $4$ consists of metagenomic families for which at least one SAE feature achieves PR-AUC$>0.5$, and since each family contains multiple constituent sequences, we report the total number of sequences in the last column.
For these families (in column $4$), high metagenomic PR-AUC suggests that the geometric annotation transfers beyond SwissProt.

\begin{table}[hbt!]          
      \centering
    \resizebox{\textwidth}{!}{%
      \begin{tabular}{c|c|c|c|c|c}                                                                                                                                                      
           & \% NMPFam act. & \% geom.~$q$-sig. & \% feat.~med.~PR-AUC $>$ 0.5 & NMPFams matched & Sequences annotated \\
           \hline                                                                                                                                                                       
           Layer 2 & 63.55 & 91.72 & 1.65\% (169) & 4.74\% (2{,}369)  & 4.40\% (440{,}475) \\                                                                                           
           Layer 4 & 77.78 & 93.50 & 3.67\% (376) & 7.75\% (3{,}875)  & 7.58\% (757{,}802) \\                                                                                           
           Layer 6 & 90.95 & 96.18 & 4.10\% (420) & 16.22\% (8{,}108) & 15.88\% (1{,}588{,}446) \\                                                                                      
      \end{tabular}                                        }                                                                                                               
      \caption{\textbf{Geometric annotation of metagenomic proteins.} Col.~1: \% of features with NMPFam activation. Col.~2: \% of   
  NMPFam activated features that are geometrically $q$-significant. Col.~3: \% of those features with median PR-AUC$>0.5$ evaluated on activating metagenomic sequences. Col.~4 \% of 50k \emph{families} with PR-AUC$>0.5$. Col.~5: \% of 10M \emph{proteins} with PR-AUC$>0.5$.}
      \label{tab:PR-AUC-metagenomic}
\end{table}

\begin{figure}[hbt!]
    \centering
    \includegraphics[width=0.8\linewidth]{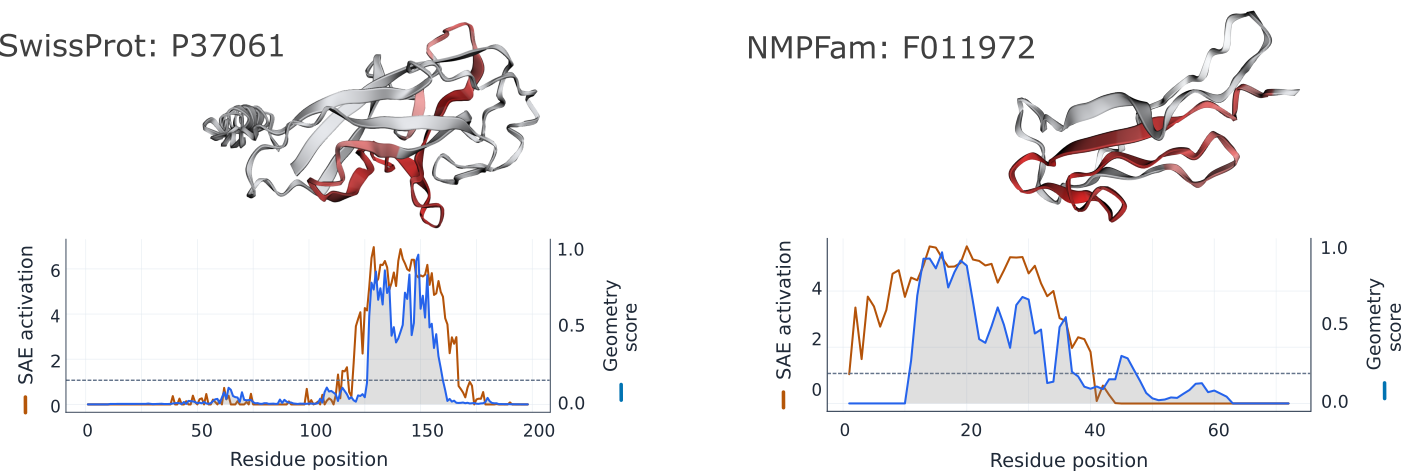}
    \caption{\textbf{Local geometric annotations transfer to metagenomic proteins.} Geometry predicts activation of a sample SwissProt protein $P37016$ well (left) and comparable perfomance on unseen metagenomic protein $F011972$ (right) with similar twisted beta sheet motifs consistently highlighted.}
    \label{fig:metagenomics}
\end{figure}

\autoref{fig:metagenomics} shows a representative transfer example for the layer $r$ feature $\f{8518}$. This feature is geometrically significant on SwissProt, with PR-AUC $0.62$, and also activates on $6$ metagenomic protein families with an average PR-AUC $0.73$. On reference SwissProt proteins, $\f{8518}$ activates on $\beta$-sheet rich regions between residues 125-175, exhibiting high twist and specific local compaction; in the metagenomic example, we observe a similar motif of two $\beta$ rich regions connected by a turn.

\textbf{4. Contact prediction response to SAE feature ablations.}
Finally, as a preliminary intervention experiment, we ask whether geometry-annotated SAE features are coupled to ESM-2 8M's structural predictions. Because ESM-2 8M does not have a 'folding head' mechanism to predict structure, we use its internal contact-map predictions as the available structural readout. For selected SAE features, we consider continuous ablation of SAE features at layer 4 (that is, we ablate on the level of the SAE feature and decode back to the `ablated' ESM-2 layer input), and subsequently analyze the resulting contact-map predictions available within ESM-2 8M.

\begin{figure}[hbt!]
    \centering
    \includegraphics[width=1\linewidth]{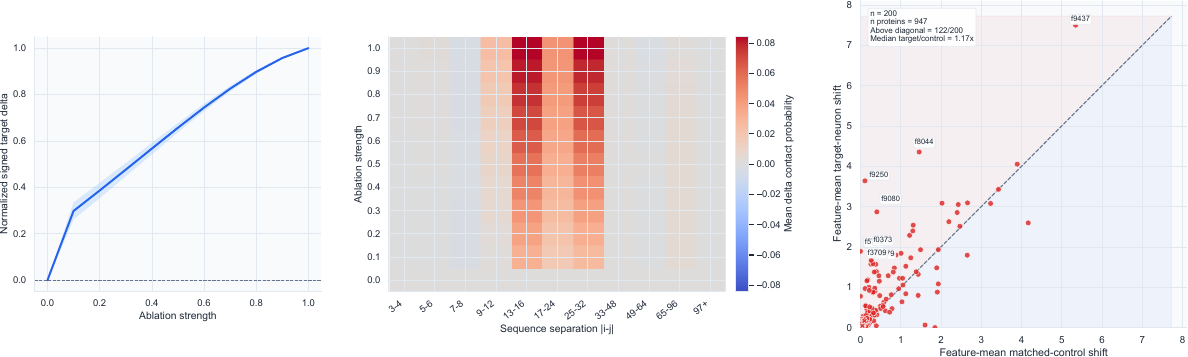}
    \caption{\textbf{Contact prediction ablation experiments at layer 4.} (Left) Ablation strength versus normalized signed target delta for SAE feature $\f{373}$, revealing the linear relationship between response in dominant geometric feature vector features and ablation strength suggesting a causal relationship. (Middle) Contact probability at binned sequence separations versus ablation strength, revealing a clear, increasing relationship between probability of contacts at sequence separations [25-48] with increasing ablation strength averaged across 20 proteins firing at $\f{373}$. (Right) Ablation experiments across 947 proteins covering the top 200 SAE features; a significant proportion see target geometric feature shift beyond matched-control tests.}
    \label{fig:ablations}
\end{figure}

In~\autoref{fig:ablations}, we first examine (left) a localized intervention on $\f{373}$, a geometry-primary, contact-related feature whose strongest geometric predictor is `mean sequence separation at 12\AA' (\autoref{sec:appendix-contacts}). Across its top 20 activating proteins, we ablate $\f{373}$ at 11 strengths (from no ablation to full ablation) and measure the induced change in the corresponding contact-geometry metric. The normalized signed target delta (the normalized signed change in the target contact-geometry metric after ablating $\f{373}$) varies consistently with ablation strength, indicating the feature is linked to the predicted contact geometry. In addition, (~\autoref{fig:ablations} middle) reveals the averaged effect of ablation on the contact probability (at <12\AA) at chosen windows across the contact map. We observe that ablation most strongly decreases predicted contacts at localized intermediate sequence separations, in particular bins $[13,48]$.
A larger ablation experiment is considered in~\autoref{fig:ablations} (right) showing significant response in the dominant geometric features (sequence separation/long range contact etc.) with respect to SAE feature across 947 proteins at 200 features in comparison to matched control-shift response. Here, the ablated feature activation is set to zero, whereas the `matched-control' considers SAE features whose activation presented similar sparsity, however varied by target vector.

\section{Discussion}

\textbf{Limitations.}
Our analysis is limited by both model scale (due to compute access) and the use of predicted protein structures. For computational reasons, we focus on ESM-2 8M, which allows for contact-map interventions but not validation through a full folding head; thus our ablations should be interpreted as contact-level interventions. There is also potential circularity in using AlphaFold-predicted geometry to annotate pLM features.
However, \cite{rao_transformer_2020} find that the high dimensional latent space of pLMs correlate significantly with contact maps of structures, which suggests hidden structure being encoded which we seek to probe. See also~\autoref{sec:appendix-real_protein} for a small validation set with experimentally determined protein structures. In addition, analysis of the geometry of metagenomic proteins is a less robust source of information as the predicted structures have a lower pLDDT. Our BH-corrected tests control the expected false-discovery proportion, but do not provide simultaneous validation of the aggregate results. Future work could address this using higher-resolution permutation tests or confirmatory aggregate-level testing.

\textbf{Outlook.}
The geometric annotations developed here suggest potential downstream applications in protein identification and engineering. While developing these applications is beyond the scope of this work, we outline two directions where our annotations can be used for future  workflows.
\begin{enumerate}
    \item\label{item:1} AlphaFold is limited in giving information about novel amino acid sequences of proteins with minimal similarity to known proteins, as its inherent mechanism relies on MSA. Hence, predictions about the structure of this new protein are limited. In contrast, pLMs do not rely on such a mechanism and as we show, the sparse geometric features found by the SAE, trained on internal representations without a prior folded structure,
    provides a coarse set of similar proteins for different geometric motifs. This can in turn provide a biologist with a starting point for comparing homologous proteins, an important aspect in prioritizing which sequences are worth pursuing for drug development/research purposes.
    
    \item Recombination, the process of sequence exchange between two homologous proteins to create a chimera with novel and desired characteristics, requires knowledge of a natural ‘seam’ along the protein~\cite{voigt_protein_2002}. In the paper, we consider the explicit case study of Q15120 and how InterPro domain labels can be resolved at finer detail using the geometric annotation methods. We argue that although these may not be characterised as the ‘seams’ themselves, biologists may be able to view them as natural candidates for potential seam positions. Well established tools for doing this like SCHEMA require the proteins 3D coordinates beforehand, and so for novel, highly unique proteins, the same limitations as in~\autoref{item:1} appear.
\end{enumerate}

\textbf{Conclusion.}
We introduce a scalable residue-level pipeline for pLM SAE features centered on geometrically-informed features. Beyond expanding annotation coverage in ESM-2 8M, these annotations provide a concrete framework for probing structure-related representations: they refine coarse biological labels into local geometric motifs, transfer to some metagenomic proteins, and are coupled to contact-map predictions under targeted ablation. 
Taken together, this work provides a step toward mechanistic interpretability for pLMs, and suggests further directions such as scaling geometric SAE annotations to larger structure-aware pLMs, validating feature interventions at the full structural level, and exploring the possibility of geometric steering for protein design.

\begin{ack}
    The authors would like to thank Sjoerd Beentjes for helpful discussions regarding the statistical methodology of the paper, including the use of permutation tests for null testing and the use of Benjamini-Hochberg corrections.
    We further thank Emma Tomlinson and Tadas Kluonis for useful discussions on interpreting our findings.
    Siddharth Setlur and Djordje Mihajlovic's work was supported by the UKRI Centre for Doctoral Training in Algebra, Geometry and Quantum Fields (AGQ), Grant Number EP/Y035232/1. 
    This work was supported by the Edinburgh International Data Facility (EIDF) and the Data-Driven Innovation Programme at the University of Edinburgh.
\end{ack}

\bibliographystyle{unsrtnat}
\bibliography{references}

\begin{thebibliography}{46}
\providecommand{\natexlab}[1]{#1}
\providecommand{\url}[1]{\texttt{#1}}
\expandafter\ifx\csname urlstyle\endcsname\relax
  \providecommand{\doi}[1]{doi: #1}\else
  \providecommand{\doi}{doi: \begingroup \urlstyle{rm}\Url}\fi

\bibitem[Vaswani et~al.(2017)Vaswani, Shazeer, Parmar, Uszkoreit, Jones, Gomez, Kaiser, and Polosukhin]{vaswani_attention_2017}
Ashish Vaswani, Noam Shazeer, Niki Parmar, Jakob Uszkoreit, Llion Jones, Aidan~N. Gomez, Łukasz Kaiser, and Illia Polosukhin.
\newblock Attention is all you need.
\newblock In \emph{Proceedings of the 31st {International} {Conference} on {Neural} {Information} {Processing} {Systems}}, {NIPS}'17, pages 6000--6010, Red Hook, NY, USA, 2017. Curran Associates Inc.
\newblock ISBN 9781510860964.
\newblock event-place: Long Beach, California, USA.

\bibitem[Ahmad et~al.(2025)Ahmad, Jose da Costa Gonzales, Bowler-Barnett, Rice, Kim, Wijerathne, Luciani, Kandasaamy, Luo, Watkins, Turner, Martin, {the UniProt Consortium}, Bateman, Martin, Orchard, Magrane, Ye, Adesina, Ahmad, Bowler-Barnett, Carpentier, Denny, Fan, Da~Costa~Gonzales, Hussein, Ignatchenko, Insana, Ishtiaq, Joshi, Jyothi, Kandasaamy, Lock, Luciani, Luo, Lussi, Marin, Raposo, Rice, Stephenson, Totoo, Urakova, Vasudev, Wijerathne, Ibrahim, Kim, Yu, Bridge, Aimo, Argoud-Puy, Auchincloss, Axelsen, Bansal, Baratin, Batista~Neto, Blatter, Bolleman, Boutet, Breuza, Casals-Casas, Echioukh, Coudert, Cuche, De~Castro, Estreicher, Famiglietti, Feuermann, Gasteiger, Gaudet, Gehant, Gerritsen, Gos, Gruaz, Hulo, Hyka-Nouspikel, Jungo, Kerhornou, Mercier, Lieberherr, Masson, Morgat, Paesano, Pedruzzi, Pilbout, Pourcel, Poux, Pozzato, Pruess, Redaschi, Rivoire, Sigrist, Sonesson, Sundaram, Sveshnikova, Wu, Chen, Huang, Laiho, Lehvaslaiho, McGarvey, Natale, Ross, Vinayaka, Wang, Mazumber, Shanker, and Zhang]{ahmad_uniprot_2025}
Shadab Ahmad, Leonardo Jose da Costa Gonzales, Emily H Bowler-Barnett, Daniel L Rice, Minjoon Kim, Supun Wijerathne, Aurélien Luciani, Swaathi Kandasaamy, Jie Luo, Xavier Watkins, Edd Turner, Maria J Martin, {the UniProt Consortium}, Alex Bateman, Maria-Jesus Martin, Sandra Orchard, Michele Magrane, Shiqi Ye, Aduragbemi Adesina, Shadab Ahmad, Emily~H Bowler-Barnett, David Carpentier, Paul Denny, Jun Fan, Leonardo~Jose Da~Costa~Gonzales, Abdulrahman Hussein, Alexandr Ignatchenko, Giuseppe Insana, Rizwan Ishtiaq, Vishal Joshi, Dushyanth Jyothi, Swaathi Kandasaamy, Antonia Lock, Aurelien Luciani, Jie Luo, Yvonne Lussi, Juan Sebastian~Martinez Marin, Pedro Raposo, Daniel~L Rice, James Stephenson, Prabhat Totoo, Nadya Urakova, Preethi Vasudev, Supun Wijerathne, Khawaja~Talal Ibrahim, Minjoon Kim, Conny Wing-Heng Yu, Alan~J Bridge, Lucila Aimo, Ghislaine Argoud-Puy, Andrea~H Auchincloss, Kristian~B Axelsen, Parit Bansal, Delphine Baratin, Teresa~M Batista~Neto, Marie-Claude Blatter, Jerven~T Bolleman, Emmanuel Boutet, Lionel Breuza, Cristina Casals-Casas, Kamal~Chikh Echioukh, Elisabeth Coudert, Beatrice Cuche, Edouard De~Castro, Anne Estreicher, Maria~L Famiglietti, Marc Feuermann, Elisabeth Gasteiger, Pascale Gaudet, Sebastien Gehant, Vivienne Gerritsen, Arnaud Gos, Nadine Gruaz, Chantal Hulo, Nevila Hyka-Nouspikel, Florence Jungo, Arnaud Kerhornou, Philippe~Le Mercier, Damien Lieberherr, Patrick Masson, Anne Morgat, Salvo Paesano, Ivo Pedruzzi, Sandrine Pilbout, Lucille Pourcel, Sylvain Poux, Monica Pozzato, Manuela Pruess, Nicole Redaschi, Catherine Rivoire, Christian J~A Sigrist, Karin Sonesson, Shyamala Sundaram, Anastasia Sveshnikova, Cathy~H Wu, Chuming Chen, Hongzhan Huang, Kati Laiho, Minna Lehvaslaiho, Peter McGarvey, Darren~A Natale, Karen Ross, C~R Vinayaka, Yuqi Wang, Raja Mazumber, Vijay Shanker, and Jian Zhang.
\newblock The {UniProt} website {API}: facilitating programmatic access to protein knowledge.
\newblock \emph{Nucleic Acids Research}, 53\penalty0 (W1):\penalty0 W547--W553, July 2025.
\newblock ISSN 0305-1048, 1362-4962.
\newblock \doi{10.1093/nar/gkaf394}.
\newblock URL \url{https://academic.oup.com/nar/article/53/W1/W547/8126256}.

\bibitem[Lin et~al.(2023)Lin, Akin, Rao, Hie, Zhu, Lu, Smetanin, Verkuil, Kabeli, Shmueli, Dos Santos~Costa, Fazel-Zarandi, Sercu, Candido, and Rives]{lin_evolutionary-scale_2023}
Zeming Lin, Halil Akin, Roshan Rao, Brian Hie, Zhongkai Zhu, Wenting Lu, Nikita Smetanin, Robert Verkuil, Ori Kabeli, Yaniv Shmueli, Allan Dos Santos~Costa, Maryam Fazel-Zarandi, Tom Sercu, Salvatore Candido, and Alexander Rives.
\newblock Evolutionary-scale prediction of atomic-level protein structure with a language model.
\newblock \emph{Science}, 379\penalty0 (6637):\penalty0 1123--1130, March 2023.
\newblock ISSN 0036-8075, 1095-9203.
\newblock \doi{10.1126/science.ade2574}.
\newblock URL \url{https://www.science.org/doi/10.1126/science.ade2574}.

\bibitem[Chen et~al.(2025)Chen, Wang, Hu, Li, Qian, and Song]{chen_evaluating_2025}
Jia-Ying Chen, Jing-Fu Wang, Yue Hu, Xin-Hui Li, Yu-Rong Qian, and Chao-Lin Song.
\newblock Evaluating the advancements in protein language models for encoding strategies in protein function prediction: a comprehensive review.
\newblock \emph{Frontiers in Bioengineering and Biotechnology}, 13:\penalty0 1506508, January 2025.
\newblock ISSN 2296-4185.
\newblock \doi{10.3389/fbioe.2025.1506508}.
\newblock URL \url{https://www.frontiersin.org/articles/10.3389/fbioe.2025.1506508/full}.

\bibitem[Jumper et~al.(2021)Jumper, Evans, Pritzel, Green, Figurnov, Ronneberger, Tunyasuvunakool, Bates, Žídek, Potapenko, Bridgland, Meyer, Kohl, Ballard, Cowie, Romera-Paredes, Nikolov, Jain, Adler, Back, Petersen, Reiman, Clancy, Zielinski, Steinegger, Pacholska, Berghammer, Bodenstein, Silver, Vinyals, Senior, Kavukcuoglu, Kohli, and Hassabis]{jumper_highly_2021}
John Jumper, Richard Evans, Alexander Pritzel, Tim Green, Michael Figurnov, Olaf Ronneberger, Kathryn Tunyasuvunakool, Russ Bates, Augustin Žídek, Anna Potapenko, Alex Bridgland, Clemens Meyer, Simon A.~A. Kohl, Andrew~J. Ballard, Andrew Cowie, Bernardino Romera-Paredes, Stanislav Nikolov, Rishub Jain, Jonas Adler, Trevor Back, Stig Petersen, David Reiman, Ellen Clancy, Michal Zielinski, Martin Steinegger, Michalina Pacholska, Tamas Berghammer, Sebastian Bodenstein, David Silver, Oriol Vinyals, Andrew~W. Senior, Koray Kavukcuoglu, Pushmeet Kohli, and Demis Hassabis.
\newblock Highly accurate protein structure prediction with {AlphaFold}.
\newblock \emph{Nature}, 596\penalty0 (7873):\penalty0 583--589, August 2021.
\newblock ISSN 0028-0836, 1476-4687.
\newblock \doi{10.1038/s41586-021-03819-2}.
\newblock URL \url{https://www.nature.com/articles/s41586-021-03819-2}.

\bibitem[Flamholz et~al.(2024)Flamholz, Biller, and Kelly]{flamholz_large_2024}
Zachary~N. Flamholz, Steven~J. Biller, and Libusha Kelly.
\newblock Large language models improve annotation of prokaryotic viral proteins.
\newblock \emph{Nature Microbiology}, 9\penalty0 (2):\penalty0 537--549, January 2024.
\newblock ISSN 2058-5276.
\newblock \doi{10.1038/s41564-023-01584-8}.
\newblock URL \url{https://www.nature.com/articles/s41564-023-01584-8}.

\bibitem[Thurimella et~al.(2025)Thurimella, Mohamed, Li, Vatanen, Graham, Owens, La~Rosa, Plichta, Bacallado, and Xavier]{thurimella_protein_2025}
Kumar Thurimella, Ahmed M.~T. Mohamed, Chenhao Li, Tommi Vatanen, Daniel~B. Graham, Róisín~M. Owens, Sabina~Leanti La~Rosa, Damian~R. Plichta, Sergio Bacallado, and Ramnik~J. Xavier.
\newblock Protein language models uncover carbohydrate-active enzyme function in metagenomics.
\newblock \emph{BMC Bioinformatics}, 26\penalty0 (1):\penalty0 285, November 2025.
\newblock ISSN 1471-2105.
\newblock \doi{10.1186/s12859-025-06286-y}.
\newblock URL \url{https://bmcbioinformatics.biomedcentral.com/articles/10.1186/s12859-025-06286-y}.

\bibitem[Rives et~al.(2021)Rives, Meier, Sercu, Goyal, Lin, Liu, Guo, Ott, Zitnick, Ma, and Fergus]{rives_biological_2021}
Alexander Rives, Joshua Meier, Tom Sercu, Siddharth Goyal, Zeming Lin, Jason Liu, Demi Guo, Myle Ott, C.~Lawrence Zitnick, Jerry Ma, and Rob Fergus.
\newblock Biological structure and function emerge from scaling unsupervised learning to 250 million protein sequences.
\newblock \emph{Proceedings of the National Academy of Sciences}, 118\penalty0 (15):\penalty0 e2016239118, April 2021.
\newblock ISSN 0027-8424, 1091-6490.
\newblock \doi{10.1073/pnas.2016239118}.
\newblock URL \url{https://pnas.org/doi/full/10.1073/pnas.2016239118}.

\bibitem[Baltoumas et~al.(2024)Baltoumas, Karatzas, Liu, Ovchinnikov, Sofianatos, Chen, Kyrpides, and Pavlopoulos]{baltoumas_nmpfamsdb_2024}
Fotis~A Baltoumas, Evangelos Karatzas, Sirui Liu, Sergey Ovchinnikov, Yorgos Sofianatos, I-Min Chen, Nikos C Kyrpides, and Georgios A Pavlopoulos.
\newblock {NMPFamsDB}: a database of novel protein families from microbial metagenomes and metatranscriptomes.
\newblock \emph{Nucleic Acids Research}, 52\penalty0 (D1):\penalty0 D502--D512, January 2024.
\newblock ISSN 0305-1048, 1362-4962.
\newblock \doi{10.1093/nar/gkad800}.
\newblock URL \url{https://academic.oup.com/nar/article/52/D1/D502/7301280}.

\bibitem[Pavlopoulos et~al.(2023)Pavlopoulos, Baltoumas, Liu, Selvitopi, Camargo, Nayfach, Azad, Roux, Call, Ivanova, Chen, Paez-Espino, Karatzas, {Novel Metagenome Protein Families Consortium}, Acinas, Ahlgren, Attwood, Baldrian, Berry, Bhatnagar, Bhaya, Bidle, Blanchard, Boyd, Bowen, Bowman, Brawley, Brodie, Brune, Bryant, Buchan, Cadillo-Quiroz, Campbell, Cavicchioli, Chuckran, Coleman, Crowe, Colman, Currie, Dangl, Delherbe, Denef, Dijkstra, Distel, Eloe-Fadrosh, Fisher, Francis, Garoutte, Gaudin, Gerwick, Godoy-Vitorino, Guerra, Guo, Habteselassie, Hallam, Hatzenpichler, Hentschel, Hess, Hirsch, Hug, Hultman, Hunt, Huntemann, Inskeep, James, Jansson, Johnston, Kalyuzhnaya, Kelly, Kelly, Klassen, Nüsslein, Kostka, Lindow, Lilleskov, Lynes, Mackelprang, Martin, Mason, McKay, McMahon, Mead, Medina, Meredith, Mock, Mohn, Moran, Murray, Neufeld, Neumann, Norton, Partida-Martinez, Pietrasiak, Pelletier, Reddy, Reese, Reichart, Reiss, Saito, Schachtman, Seshadri, Shade, Sherman, Simister, Simon, Stegen, Stepanauskas, Sullivan, Sumner, Teeling, Thamatrakoln, Treseder, Tringe, Vaishampayan, Valentine, Waldo, Waldrop, Walsh, Ward, Wilkins, Whitman, Woolet, Woyke, Iliopoulos, Konstantinidis, Tiedje, Pett-Ridge, Baker, Visel, Ouzounis, Ovchinnikov, Buluç, and Kyrpides]{pavlopoulos_unraveling_2023}
Georgios~A. Pavlopoulos, Fotis~A. Baltoumas, Sirui Liu, Oguz Selvitopi, Antonio~Pedro Camargo, Stephen Nayfach, Ariful Azad, Simon Roux, Lee Call, Natalia~N. Ivanova, I.~Min Chen, David Paez-Espino, Evangelos Karatzas, {Novel Metagenome Protein Families Consortium}, Silvia~G. Acinas, Nathan Ahlgren, Graeme Attwood, Petr Baldrian, Timothy Berry, Jennifer~M. Bhatnagar, Devaki Bhaya, Kay~D. Bidle, Jeffrey~L. Blanchard, Eric~S. Boyd, Jennifer~L. Bowen, Jeff Bowman, Susan~H. Brawley, Eoin~L. Brodie, Andreas Brune, Donald~A. Bryant, Alison Buchan, Hinsby Cadillo-Quiroz, Barbara~J. Campbell, Ricardo Cavicchioli, Peter~F. Chuckran, Maureen Coleman, Sean Crowe, Daniel~R. Colman, Cameron~R. Currie, Jeff Dangl, Nathalie Delherbe, Vincent~J. Denef, Paul Dijkstra, Daniel~D. Distel, Emiley Eloe-Fadrosh, Kirsten Fisher, Christopher Francis, Aaron Garoutte, Amelie Gaudin, Lena Gerwick, Filipa Godoy-Vitorino, Peter Guerra, Jiarong Guo, Mussie~Y. Habteselassie, Steven~J. Hallam, Roland Hatzenpichler, Ute Hentschel, Matthias Hess, Ann~M. Hirsch, Laura~A. Hug, Jenni Hultman, Dana~E. Hunt, Marcel Huntemann, William~P. Inskeep, Timothy~Y. James, Janet Jansson, Eric~R. Johnston, Marina Kalyuzhnaya, Charlene~N. Kelly, Robert~M. Kelly, Jonathan~L. Klassen, Klaus Nüsslein, Joel~E. Kostka, Steven Lindow, Erik Lilleskov, Mackenzie Lynes, Rachel Mackelprang, Francis~M. Martin, Olivia~U. Mason, R.~Michael McKay, Katherine McMahon, David~A. Mead, Monica Medina, Laura~K. Meredith, Thomas Mock, William~W. Mohn, Mary~Ann Moran, Alison Murray, Josh~D. Neufeld, Rebecca Neumann, Jeanette~M. Norton, Laila~P. Partida-Martinez, Nicole Pietrasiak, Dale Pelletier, T.~B.~K. Reddy, Brandi~Kiel Reese, Nicholas~J. Reichart, Rebecca Reiss, Mak~A. Saito, Daniel~P. Schachtman, Rekha Seshadri, Ashley Shade, David Sherman, Rachel Simister, Holly Simon, James Stegen, Ramunas Stepanauskas, Matthew Sullivan, Dawn~Y. Sumner, Hanno Teeling, Kimberlee Thamatrakoln, Kathleen Treseder, Susannah Tringe, Parag Vaishampayan, David~L. Valentine, Nicholas~B. Waldo, Mark~P. Waldrop, David~A. Walsh, David~M. Ward, Michael Wilkins, Thea Whitman, Jamie Woolet, Tanja Woyke, Ioannis Iliopoulos, Konstantinos Konstantinidis, James~M. Tiedje, Jennifer Pett-Ridge, David Baker, Axel Visel, Christos~A. Ouzounis, Sergey Ovchinnikov, Aydin Buluç, and Nikos~C. Kyrpides.
\newblock Unraveling the functional dark matter through global metagenomics.
\newblock \emph{Nature}, 622\penalty0 (7983):\penalty0 594--602, October 2023.
\newblock ISSN 0028-0836, 1476-4687.
\newblock \doi{10.1038/s41586-023-06583-7}.
\newblock URL \url{https://www.nature.com/articles/s41586-023-06583-7}.

\bibitem[Brandes et~al.(2023)Brandes, Goldman, Wang, Ye, and Ntranos]{brandes_genome-wide_2023}
Nadav Brandes, Grant Goldman, Charlotte~H. Wang, Chun~Jimmie Ye, and Vasilis Ntranos.
\newblock Genome-wide prediction of disease variant effects with a deep protein language model.
\newblock \emph{Nature Genetics}, 55\penalty0 (9):\penalty0 1512--1522, September 2023.
\newblock ISSN 1061-4036, 1546-1718.
\newblock \doi{10.1038/s41588-023-01465-0}.
\newblock URL \url{https://www.nature.com/articles/s41588-023-01465-0}.

\bibitem[Ismail et~al.(2024)Ismail, Oikarinen, Wang, Adebayo, Stanton, Joren, Kleinhenz, Goodman, Bravo, Cho, and Frey]{ismail_concept_2024}
Aya~Abdelsalam Ismail, Tuomas Oikarinen, Amy Wang, Julius Adebayo, Samuel Stanton, Taylor Joren, Joseph Kleinhenz, Allen Goodman, Héctor~Corrada Bravo, Kyunghyun Cho, and Nathan~C. Frey.
\newblock Concept {Bottleneck} {Language} {Models} {For} protein design, November 2024.
\newblock URL \url{https://arxiv.org/abs/2411.06090v2}.

\bibitem[Olah et~al.(2020)Olah, Cammarata, Schubert, Goh, Petrov, and Carter]{olah_zoom_2020}
Chris Olah, Nick Cammarata, Ludwig Schubert, Gabriel Goh, Michael Petrov, and Shan Carter.
\newblock Zoom {In}: {An} {Introduction} to {Circuits}.
\newblock \emph{Distill}, 5\penalty0 (3):\penalty0 10.23915/distill.00024.001, March 2020.
\newblock ISSN 2476-0757.
\newblock \doi{10.23915/distill.00024.001}.
\newblock URL \url{https://distill.pub/2020/circuits/zoom-in}.

\bibitem[Elhage et~al.(2022)Elhage, Hume, Olsson, Schiefer, Henighan, Kravec, Hatfield-Dodds, Lasenby, Drain, Chen, Grosse, McCandlish, Kaplan, Amodei, Wattenberg, and Olah]{elhage_toy_2022}
Nelson Elhage, Tristan Hume, Catherine Olsson, Nicholas Schiefer, Tom Henighan, Shauna Kravec, Zac Hatfield-Dodds, Robert Lasenby, Dawn Drain, Carol Chen, Roger Grosse, Sam McCandlish, Jared Kaplan, Dario Amodei, Martin Wattenberg, and Christopher Olah.
\newblock Toy {Models} of {Superposition}, 2022.
\newblock URL \url{https://arxiv.org/abs/2209.10652}.

\bibitem[Paulo et~al.(2025)Paulo, Mallen, Juang, and Belrose]{paulo_automatically_2025}
Gonçalo Paulo, Alex Mallen, Caden Juang, and Nora Belrose.
\newblock Automatically {Interpreting} {Millions} of {Features} in {Large} {Language} {Models}, August 2025.
\newblock URL \url{http://arxiv.org/abs/2410.13928}.
\newblock arXiv:2410.13928.

\bibitem[Steven et~al.(2023)Steven, Cammarata, Mossing, Tillman, Gao, Goh, Sutskever, Leike, Wu, and Saunders]{steven_language_2023}
BIll Steven, Nick Cammarata, Dan Mossing, Henk Tillman, Leo Gao, Gabriel Goh, Ilya Sutskever, Jan Leike, Jeff Wu, and William Saunders.
\newblock Language models can explain neurons in language models, May 2023.
\newblock URL \url{https://openaipublic.blob.core.windows.net/neuron-explainer/paper/index.html#sec-limitations}.

\bibitem[Simon and Zou(2025)]{simon_interplm_2025}
Elana Simon and James Zou.
\newblock {InterPLM}: discovering interpretable features in protein language models via sparse autoencoders.
\newblock \emph{Nature Methods}, 22\penalty0 (10):\penalty0 2107--2117, October 2025.
\newblock ISSN 1548-7091, 1548-7105.
\newblock \doi{10.1038/s41592-025-02836-7}.
\newblock URL \url{https://www.nature.com/articles/s41592-025-02836-7}.

\bibitem[Adams et~al.(2025)Adams, Bai, Lee, Yu, and AlQuraishi]{adams_mechanistic_2025}
Etowah Adams, Liam Bai, Minji Lee, Yiyang Yu, and Mohammed AlQuraishi.
\newblock From {Mechanistic} {Interpretability} to {Mechanistic} {Biology}: {Training}, {Evaluating}, and {Interpreting} {Sparse} {Autoencoders} on {Protein} {Language} {Models}.
\newblock \emph{bioRxiv: The Preprint Server for Biology}, page 2025.02.06.636901, June 2025.
\newblock ISSN 2692-8205.
\newblock \doi{10.1101/2025.02.06.636901}.

\bibitem[Silberg et~al.(2025)Silberg, Simon, and Zou]{silberg_towards_2025}
Jake Silberg, Elana Simon, and James Zou.
\newblock Towards functional annotation with latent protein language model features, October 2025.
\newblock URL \url{https://www.biorxiv.org/content/10.1101/2025.10.02.680154v1}.

\bibitem[Blum et~al.(2025)Blum, Andreeva, Florentino, Chuguransky, Grego, Hobbs, Pinto, Orr, Paysan-Lafosse, Ponamareva, Salazar, Bordin, Bork, Bridge, Colwell, Gough, Haft, Letunic, Llinares-López, Marchler-Bauer, Meng-Papaxanthos, Mi, Natale, Orengo, Pandurangan, Piovesan, Rivoire, Sigrist, Thanki, Thibaud-Nissen, Thomas, Tosatto, Wu, and Bateman]{blum_interpro_2025}
Matthias Blum, Antonina Andreeva, Laise Cavalcanti Florentino, Sara Rocio Chuguransky, Tiago Grego, Emma Hobbs, Beatriz Lazaro Pinto, Ailsa Orr, Typhaine Paysan-Lafosse, Irina Ponamareva, Gustavo A Salazar, Nicola Bordin, Peer Bork, Alan Bridge, Lucy Colwell, Julian Gough, Daniel H Haft, Ivica Letunic, Felipe Llinares-López, Aron Marchler-Bauer, Laetitia Meng-Papaxanthos, Huaiyu Mi, Darren A Natale, Christine A Orengo, Arun P Pandurangan, Damiano Piovesan, Catherine Rivoire, Christian J~A Sigrist, Narmada Thanki, Françoise Thibaud-Nissen, Paul D Thomas, Silvio C~E Tosatto, Cathy H Wu, and Alex Bateman.
\newblock {InterPro}: the protein sequence classification resource in 2025.
\newblock \emph{Nucleic Acids Research}, 53\penalty0 (D1):\penalty0 D444--D456, January 2025.
\newblock ISSN 0305-1048, 1362-4962.
\newblock \doi{10.1093/nar/gkae1082}.
\newblock URL \url{https://academic.oup.com/nar/article/53/D1/D444/7905301}.

\bibitem[Paysan-Lafosse et~al.(2025)Paysan-Lafosse, Andreeva, Blum, Chuguransky, Grego, Pinto, Salazar, Bileschi, Llinares-López, Meng-Papaxanthos, Colwell, Grishin, Schaeffer, Clementel, Tosatto, Sonnhammer, Wood, and Bateman]{paysan-lafosse_pfam_2025}
Typhaine Paysan-Lafosse, Antonina Andreeva, Matthias Blum, Sara Rocio Chuguransky, Tiago Grego, Beatriz Lazaro Pinto, Gustavo A Salazar, Maxwell L Bileschi, Felipe Llinares-López, Laetitia Meng-Papaxanthos, Lucy J Colwell, Nick V Grishin, R~Dustin Schaeffer, Damiano Clementel, Silvio C~E Tosatto, Erik Sonnhammer, Valerie Wood, and Alex Bateman.
\newblock The {Pfam} protein families database: embracing {AI}/{ML}.
\newblock \emph{Nucleic Acids Research}, 53\penalty0 (D1):\penalty0 D523--D534, January 2025.
\newblock ISSN 0305-1048, 1362-4962.
\newblock \doi{10.1093/nar/gkae997}.
\newblock URL \url{https://academic.oup.com/nar/article/53/D1/D523/7900195}.

\bibitem[Sillitoe et~al.(2021)Sillitoe, Bordin, Dawson, Waman, Ashford, Scholes, Pang, Woodridge, Rauer, Sen, Abbasian, Le Cornu, Lam, Berka, Varekova, Svobodova, Lees, and Orengo]{sillitoe_cath_2021}
Ian Sillitoe, Nicola Bordin, Natalie Dawson, Vaishali~P Waman, Paul Ashford, Harry~M Scholes, Camilla S~M Pang, Laurel Woodridge, Clemens Rauer, Neeladri Sen, Mahnaz Abbasian, Sean Le Cornu, Su~Datt Lam, Karel Berka, Ivana Hutařová Varekova, Radka Svobodova, Jon Lees, and Christine~A Orengo.
\newblock {CATH}: increased structural coverage of functional space.
\newblock \emph{Nucleic Acids Research}, 49\penalty0 (D1):\penalty0 D266--D273, January 2021.
\newblock ISSN 0305-1048, 1362-4962.
\newblock \doi{10.1093/nar/gkaa1079}.
\newblock URL \url{https://academic.oup.com/nar/article/49/D1/D266/6006195}.

\bibitem[Bailey et~al.(2015)Bailey, Johnson, Grant, and Noble]{bailey_meme_2015}
Timothy~L. Bailey, James Johnson, Charles~E. Grant, and William~S. Noble.
\newblock The {MEME} {Suite}.
\newblock \emph{Nucleic Acids Research}, 43\penalty0 (W1):\penalty0 W39--W49, July 2015.
\newblock ISSN 0305-1048, 1362-4962.
\newblock \doi{10.1093/nar/gkv416}.
\newblock URL \url{https://academic.oup.com/nar/article-lookup/doi/10.1093/nar/gkv416}.

\bibitem[Rao et~al.(2020)Rao, Meier, Sercu, Ovchinnikov, and Rives]{rao_transformer_2020}
Roshan Rao, Joshua Meier, Tom Sercu, Sergey Ovchinnikov, and Alexander Rives.
\newblock Transformer protein language models are unsupervised structure learners, December 2020.
\newblock URL \url{http://biorxiv.org/lookup/doi/10.1101/2020.12.15.422761}.

\bibitem[Vig et~al.(2020)Vig, Madani, Varshney, Xiong, Socher, and Rajani]{vig_bertology_2020}
Jesse Vig, Ali Madani, Lav~R. Varshney, Caiming Xiong, Richard Socher, and Nazneen~Fatema Rajani.
\newblock {BERTology} {Meets} {Biology}: {Interpreting} {Attention} in {Protein} {Language} {Models}, 2020.
\newblock URL \url{https://arxiv.org/abs/2006.15222}.

\bibitem[Valeriani et~al.(2023)Valeriani, Doimo, Cuturello, Laio, Ansuini, and Cazzaniga]{valeriani_geometry_2023}
Lucrezia Valeriani, Diego Doimo, Francesca Cuturello, Alessandro Laio, Alessio Ansuini, and Alberto Cazzaniga.
\newblock The geometry of hidden representations of large transformer models, October 2023.
\newblock URL \url{http://arxiv.org/abs/2302.00294}.
\newblock arXiv:2302.00294.

\bibitem[Garcia and Ansuini(2025)]{garcia_interpreting_2025}
Edith Natalia~Villegas Garcia and Alessio Ansuini.
\newblock Interpreting and {Steering} {Protein} {Language} {Models} through {Sparse} {Autoencoders}, February 2025.
\newblock URL \url{http://arxiv.org/abs/2502.09135}.
\newblock arXiv:2502.09135.

\bibitem[Parsan et~al.(2025)Parsan, Yang, and Yang]{parsan_towards_2025}
Nithin Parsan, David~J. Yang, and John~J. Yang.
\newblock Towards {Interpretable} {Protein} {Structure} {Prediction} with {Sparse} {Autoencoders}, March 2025.
\newblock URL \url{http://arxiv.org/abs/2503.08764}.
\newblock arXiv:2503.08764.

\bibitem[Liu et~al.(2026)Liu, Lei, Liu, Liu, and Hu]{liu_protsae_2026}
Xiangyu Liu, Haodi Lei, Yi~Liu, Yang Liu, and Wei Hu.
\newblock {ProtSAE}: {Disentangling} and {Interpreting} {Protein} {Language} {Models} via {Semantically}-{Guided} {Sparse} {Autoencoders}, January 2026.
\newblock URL \url{http://arxiv.org/abs/2509.05309}.
\newblock arXiv:2509.05309.

\bibitem[Gujral et~al.(2025)Gujral, Bafna, Alm, and Berger]{gujral_sparse_2025}
Onkar Gujral, Mihir Bafna, Eric Alm, and Bonnie Berger.
\newblock Sparse autoencoders uncover biologically interpretable features in protein language model representations.
\newblock \emph{Proceedings of the National Academy of Sciences}, 122\penalty0 (34):\penalty0 e2506316122, August 2025.
\newblock ISSN 0027-8424, 1091-6490.
\newblock \doi{10.1073/pnas.2506316122}.
\newblock URL \url{https://pnas.org/doi/10.1073/pnas.2506316122}.

\bibitem[Devlin et~al.(2019)Devlin, Chang, Lee, and Toutanova]{devlin_bert_2019}
Jacob Devlin, Ming-Wei Chang, Kenton Lee, and Kristina Toutanova.
\newblock {BERT}: {Pre}-training of {Deep} {Bidirectional} {Transformers} for {Language} {Understanding}.
\newblock In \emph{Proceedings of the 2019 {Conference} of the {North}}, pages 4171--4186, Minneapolis, Minnesota, 2019. Association for Computational Linguistics.
\newblock \doi{10.18653/v1/N19-1423}.
\newblock URL \url{http://aclweb.org/anthology/N19-1423}.

\bibitem[Bussmann et~al.(2025)Bussmann, Nabeshima, Karvonen, and Nanda]{bussmann_learning_2025}
Bart Bussmann, Noa Nabeshima, Adam Karvonen, and Neel Nanda.
\newblock Learning {Multi}-{Level} {Features} with {Matryoshka} {Sparse} {Autoencoders}.
\newblock In \emph{Proceedings of the 42nd {International} {Conference} on {Machine} {Learning}}, pages 6077--6101. PMLR, October 2025.
\newblock URL \url{https://proceedings.mlr.press/v267/bussmann25a.html}.

\bibitem[Gao et~al.(2024)Gao, Tour, Tillman, Goh, Troll, Radford, Sutskever, Leike, and Wu]{gao_scaling_2024}
Leo Gao, Tom Dupré~la Tour, Henk Tillman, Gabriel Goh, Rajan Troll, Alec Radford, Ilya Sutskever, Jan Leike, and Jeffrey Wu.
\newblock Scaling and evaluating sparse autoencoders, June 2024.
\newblock URL \url{http://arxiv.org/abs/2406.04093}.
\newblock arXiv:2406.04093.

\bibitem[Bricken et~al.(2023)Bricken, Templeton, Batson, Chen, Jermyn, Conerly, Turner, Anil, Denison, Askell, Lasenby, Wu, Kravec, Schiefer, Maxwell, Joseph, Hatfield-Dodds, Tamkin, Nguyen, McLean, Burke, Hume, Carter, Henighan, and Olah]{bricken_towards_2023}
Trenton Bricken, Adly Templeton, Joshua Batson, Brian Chen, Adam Jermyn, Tom Conerly, Nick Turner, Cem Anil, Carson Denison, Amanda Askell, Robert Lasenby, Yifan Wu, Shauna Kravec, Nicholas Schiefer, Tim Maxwell, Nicholas Joseph, Zac Hatfield-Dodds, Alex Tamkin, Karina Nguyen, Brayden McLean, Josiah~E Burke, Tristan Hume, Shan Carter, Tom Henighan, and Christopher Olah.
\newblock Towards {Monosemanticity}: {Decomposing} {Language} {Models} {With} {Dictionary} {Learning}.
\newblock \emph{Transformer Circuits Thread}, 2023.

\bibitem[Steinegger and Söding(2017)]{steinegger_mmseqs2_2017}
Martin Steinegger and Johannes Söding.
\newblock {MMseqs2} enables sensitive protein sequence searching for the analysis of massive data sets.
\newblock \emph{Nature Biotechnology}, 35\penalty0 (11):\penalty0 1026--1028, November 2017.
\newblock ISSN 1087-0156, 1546-1696.
\newblock \doi{10.1038/nbt.3988}.
\newblock URL \url{https://www.nature.com/articles/nbt.3988}.

\bibitem[Hou et~al.(2025)Hou, Liu, and Shen]{hou_motifae_2025}
Chao Hou, Di~Liu, and Yufeng Shen.
\newblock {MotifAE} {Reveals} {Functional} {Sequence} {Patterns} from {Protein} {Language} {Model}: {Unsupervised} {Discovery} and {Interpretability} {Analysis}, November 2025.
\newblock URL \url{http://biorxiv.org/lookup/doi/10.1101/2025.11.04.686576}.

\bibitem[Ramachandran et~al.(1963)Ramachandran, Ramakrishnan, and Sasisekharan]{ramachandran_stereochemistry_1963}
G.N. Ramachandran, C.~Ramakrishnan, and V.~Sasisekharan.
\newblock Stereochemistry of polypeptide chain configurations.
\newblock \emph{Journal of Molecular Biology}, 7\penalty0 (1):\penalty0 95--99, July 1963.
\newblock ISSN 00222836.
\newblock \doi{10.1016/S0022-2836(63)80023-6}.
\newblock URL \url{https://linkinghub.elsevier.com/retrieve/pii/S0022283663800236}.

\bibitem[Oldfield and Hubbard(1994)]{oldfield_analysis_1994}
T.~J. Oldfield and R.~E. Hubbard.
\newblock Analysis of {C$\alpha$} geometry in protein structures.
\newblock \emph{Proteins: Structure, Function, and Bioinformatics}, 18\penalty0 (4):\penalty0 324--337, April 1994.
\newblock ISSN 0887-3585, 1097-0134.
\newblock \doi{10.1002/prot.340180404}.
\newblock URL \url{https://onlinelibrary.wiley.com/doi/10.1002/prot.340180404}.

\bibitem[Durairaj et~al.(2020)Durairaj, Akdel, De~Ridder, and Van~Dijk]{durairaj_geometricus_2020}
Janani Durairaj, Mehmet Akdel, Dick De~Ridder, and Aalt D~J Van~Dijk.
\newblock Geometricus represents protein structures as shape-mers derived from moment invariants.
\newblock \emph{Bioinformatics}, 36\penalty0 (Supplement\_2):\penalty0 i718--i725, December 2020.
\newblock ISSN 1367-4803, 1367-4811.
\newblock \doi{10.1093/bioinformatics/btaa839}.
\newblock URL \url{https://academic.oup.com/bioinformatics/article/36/Supplement_2/i718/6055902}.

\bibitem[Wiedemann et~al.(2020)Wiedemann, Kumar, Lang, and Ohlenschläger]{wiedemann_cysteines_2020}
Christoph Wiedemann, Amit Kumar, Andras Lang, and Oliver Ohlenschläger.
\newblock Cysteines and {Disulfide} {Bonds} as {Structure}-{Forming} {Units}: {Insights} {From} {Different} {Domains} of {Life} and the {Potential} for {Characterization} by {NMR}.
\newblock \emph{Frontiers in Chemistry}, 8:\penalty0 280, April 2020.
\newblock ISSN 2296-2646.
\newblock \doi{10.3389/fchem.2020.00280}.
\newblock URL \url{https://www.frontiersin.org/article/10.3389/fchem.2020.00280/full}.

\bibitem[Schaeffer et~al.(2026)Schaeffer, Zhang, Cong, and Grishin]{schaeffer_ecod_2026}
R.~Dustin Schaeffer, Jing Zhang, Qian Cong, and Nick~V. Grishin.
\newblock {ECOD}: {Classification} of domains in {AFDB} {Swiss}-{Prot} structure predictions.
\newblock \emph{PLOS Computational Biology}, 22\penalty0 (3):\penalty0 e1013431, March 2026.
\newblock ISSN 1553-7358.
\newblock \doi{10.1371/journal.pcbi.1013431}.
\newblock URL \url{https://dx.plos.org/10.1371/journal.pcbi.1013431}.

\bibitem[Phipson and Smyth(2010)]{phipson_permutation_2010}
Belinda Phipson and Gordon~K Smyth.
\newblock Permutation {P}-values {Should} {Never} {Be} {Zero}: {Calculating} {Exact} {P}-values {When} {Permutations} {Are} {Randomly} {Drawn}.
\newblock \emph{Statistical Applications in Genetics and Molecular Biology}, 9\penalty0 (1), October 2010.
\newblock ISSN 1544-6115.
\newblock \doi{10.2202/1544-6115.1585}.
\newblock URL \url{https://www.degruyterbrill.com/document/doi/10.2202/1544-6115.1585/html}.

\bibitem[Benjamini and Hochberg(1995)]{benjamini_controlling_1995}
Yoav Benjamini and Yosef Hochberg.
\newblock Controlling the {False} {Discovery} {Rate}: {A} {Practical} and {Powerful} {Approach} to {Multiple} {Testing}.
\newblock \emph{Journal of the Royal Statistical Society Series B: Statistical Methodology}, 57\penalty0 (1):\penalty0 289--300, January 1995.
\newblock ISSN 1369-7412, 1467-9868.
\newblock \doi{10.1111/j.2517-6161.1995.tb02031.x}.
\newblock URL \url{https://academic.oup.com/jrsssb/article/57/1/289/7035855}.

\bibitem[Rego and Koes(2015)]{rego_3dmoljs_2015}
Nicholas Rego and David Koes.
\newblock {3Dmol}.js: molecular visualization with {WebGL}.
\newblock \emph{Bioinformatics}, 31\penalty0 (8):\penalty0 1322--1324, April 2015.
\newblock ISSN 1367-4811, 1367-4803.
\newblock \doi{10.1093/bioinformatics/btu829}.
\newblock URL \url{https://academic.oup.com/bioinformatics/article/31/8/1322/213186}.

\bibitem[Voigt et~al.(2002)Voigt, Martinez, Wang, Mayo, and Arnold]{voigt_protein_2002}
Christopher~A. Voigt, Carlos Martinez, Zhen-Gang Wang, Stephen~L. Mayo, and Frances~H. Arnold.
\newblock Protein building blocks preserved by recombination.
\newblock \emph{Nature Structural Biology}, June 2002.
\newblock ISSN 10728368.
\newblock \doi{10.1038/nsb805}.
\newblock URL \url{https://www.nature.com/doifinder/10.1038/nsb805}.

\bibitem[Marks et~al.(2024)Marks, Karvonen, and Mueller]{marks_dictionary_learning_2024}
Samuel Marks, Adam Karvonen, and Adam Mueller.
\newblock dictionary\_learning, 2024.
\newblock URL \url{https://github.com/saprmarks/dictionary_learning}.

\end{thebibliography}

\appendix

\clearpage
\section{SAE training details}
\label{sec:appendix-sae-training}

We acknowledge the usage of Claude Code to help with code management and visualization development.

On training and optimizing our models we use a compute cluster to access 4 Nvidia A100's. We implement ML algorithms primarily using PyTorch and SciKit-learn.

We train SAEs for layer 2,4, and 6 of the residual stream of ESM2-8M following the architecture detailed in \cite{bricken_towards_2023} and the training recipe detailed in \cite{marks_dictionary_learning_2024}. This architecutre and training recipe was used in the pLM setting by \cite{simon_interplm_2025} and we draw heavily from their codebase and training details. 
\subsection{Training data}
We train on $50,000$ uniformly sampled sequences from the SwissProt database \cite{ahmad_uniprot_2025} filtering for sequences with length $\leq 1024$ (this is the maximum length accepted by ESM2-8M), holding out one shad for fidelity evaluation. We then run each of these sequences through ESM2-8M and extract per-residue activation vectors of the residual streams at layers $2,4,6$. These $320$ dimensional activation vectors are used to train and evaluate (with the held out set) the SAEs at the respective layer.

\subsection{SAE architecture details}

We constrain each decoder dictionary vector to have unit $\ell_2$ norm. In implementation terms, because the decoder is stored as a linear layer with weight matrix of shape $d \times m$, this is enforced as a unit-norm constraint on the columns of $W_{\mathrm{dec}}$. During optimization, we project decoder gradients to be orthogonal to the current decoder directions and renormalize the decoder columns after every optimizer step. This keeps each decoder feature vector on the unit sphere throughout training.

At initialization, the shared bias $b$ is set to zero. The decoder weights are initialized from an isotropic Gaussian and then column-normalized to unit norm. Encoder weights and bias are initialized independently using the default \texttt{nn.Linear} initialization in PyTorch.

Note, the encoder and decoder are untied; that is the encoder weights and decoder weights are learned as two separate parameter matrices.

\subsubsection{Hyperparameter search}

  \begin{table}[hbt!]
  \centering
  \caption{Hyperparameter sweep configuration.}
  \label{tab:sweep_search_space}
  \small
  \begin{tabular}{@{}llll@{}}
  \toprule
  Hyperparameter & Distribution & Range / values & Notes \\
  \midrule
  Learning rate $\eta$        & log-uniform & $[5\!\times\!10^{-5},\,10^{-3}]$ &
   Adam initial LR \\
  L1 coefficient $\lambda$    & log-uniform & $[10^{-2},\,0.2]$                &
   full strength after warmup \\
  Dictionary size $m$         & categorical &
  $\{2{,}560,\,5{,}120,\,10{,}240,\,20{,}480\}$ & 8$\times$, 16$\times$,
  32$\times$, 64$\times$ \\
  Batch size                  & categorical & $\{1{,}024,\,2{,}048,\,4{,}096\}$
  & residue tokens / step \\
  Warmup ratio                & uniform     & $[0.05,\,0.15]$                  &
   of 500\,000 total steps \\
  Decay-start ratio           & uniform     & $[0.70,\,0.90]$                  &
   of 500\,000 total steps \\
  \midrule
  Search method     & \multicolumn{3}{l}{Bayesian optimisation} \\
  Objective         & \multicolumn{3}{l}{maximise
  \texttt{performance/pct\_loss\_recovered}} \\
  Early termination & \multicolumn{3}{l}{Hyperband, \texttt{min\_iter} $=10^{5}$
   steps} \\
  Runs (layer 2 / 4 / 6) & \multicolumn{3}{l}{11 / 33 / $\sim$30 completed runs
  per sweep} \\
  Selection rule    & \multicolumn{3}{l}{best fidelity at step 500\,000 among
  non-collapsed runs} \\
  \bottomrule
  \end{tabular}
  \end{table}

  \clearpage
\subsubsection{Selected SAEs}

\begin{table}[h]
  \centering
  \caption{Selected SAE per layer. All three SAEs are L1-ReLU autoencoders
  with the architecture of \cite{bricken_towards_2023} and the training
  recipe of \cite{simon_interplm_2025}.}
  \label{tab:selected_saes}
  \small
  \setlength{\tabcolsep}{6pt}
  \begin{tabular}{@{}lrrr@{}}
  \toprule
  & \textbf{Layer 2} & \textbf{Layer 4} & \textbf{Layer 6} \\
  \midrule
  \multicolumn{4}{l}{\textit{Architecture}} \\
  \quad Activation dim $d$       & 320      & 320      & 320 \\
  \quad Dictionary size $m$      & 10\,240  & 10\,240  & 10\,240 \\
  \quad Expansion factor         & 32$\times$ & 32$\times$ & 32$\times$ \\
  \midrule
  \multicolumn{4}{l}{\textit{Optimisation}} \\
  \quad Batch size               & 4\,096   & 2\,048   & 1\,024 \\
  \quad Learning rate $\eta$     & $1.890{\times}10^{-4}$ &
  $5.367{\times}10^{-5}$ & $3.209{\times}10^{-4}$ \\
  \quad L1 coefficient $\lambda$ & 0.0759   & 0.0794   & 0.0522 \\
  \quad LR warmup steps          & 41\,246  & 43\,907  & 64\,189 \\
  \quad LR decay-start step      & 386\,139 & 399\,947 & 407\,488 \\
  \quad Total steps              & 500\,000 & 500\,000 & 500\,000 \\
  \quad Epochs over training set & 170.4    & 85.2     & 42.6 \\
  \quad Tokens seen              & $2.05\!\times\!10^{9}$ &
  $1.02\!\times\!10^{9}$ & $5.12\!\times\!10^{8}$ \\
  \midrule
  \multicolumn{4}{l}{\textit{Final-step performance (held-out eval shard)}} \\
  \quad \% loss recovered        & 99.29 & 99.34 & 100.00 \\
  \quad CE w/ SAE patching       & 0.4590 & 0.4607 & 0.4336 \\
  \quad Variance explained       & 0.9195 & 0.9210 & 0.9637 \\
  \quad $\ell_0$ sparsity (active feats./token) & 335.3 & 252.3 & 726.5 \\
  \quad Reconstruction loss $\|\!\cdot\!\|_2$   & 1.974 & 4.041 & 0.891 \\
  \quad Sparsity loss $\|\!\cdot\!\|_1$         & 24.13 & 40.41 & 23.65 \\
  \quad Total loss               & 3.806 & 7.251 & 2.126 \\
  \quad Dead features ($<\!100$ steps)  & 580 & 51 & 160 \\
  \quad Dead features ($<\!1000$ steps) & 533 & 9  & 155 \\
  \bottomrule
  \end{tabular}
  \end{table}

\subsection{Computational Resources}
The following computational resources were used (with computation times provided for a single layer). 

For SAE training, we used NVIDIA A100-SXM4-80GB GPUs, with an approximate training time of 5 hours. Memory usage was 35GB VRAM mean and 5GB CPU RSS peak. For the annotation pipeline total runtime was approximate 120 hours total. This consisted of
\begin{itemize}
    \item \textbf{Surveying per feature activation:} 17 hours total using NVIDIA A100-SXM4-80GB GPU
    \item \textbf{GBM fitting and geometry residue enrichment}: 70 hours on CPU
    \item \textbf{Database/position methods:} 35 hours on CPU
\end{itemize}

The statistical tests were run on CPU, which took approximately 14 hours.

\clearpage
\section{Sampling and Thresholding for Annotation Pipeline} \label{sec:appendix-annotation-pipeline}

Here we provide further details on the annotation pipeline. 

\textbf{Sampling Strategy.}
In order to effectively annotate an SAE feature using any of our annotatoin methods, we first need a representative sample of protein sequences that the SAE fires on. We use two slightly different sampling strategies reflecting two regimes of annotation methods: 
\begin{enumerate}
    \item For methods that output a continuous score in $[0,1]$ at each residue position to predict SAE activation we sample up to $n$ of the top activating proteins at that feature. These methods require windows of length $l$ around top activating residue positions of each sequence for training models or computing probability matrices, and so we only select retain with length greater than $l$. 
    \item For methods that output a binary prediction of SAE activation at each residue position, we normalize the feature's per-protein maximum activation by its global maximum across the proteome and stratify proteins into $11$ bins: one inactive bin (activation = $0$) and ten equal-width bins $[0.0, 0.1), …, [0.9, 1.0]$. We select up to $m$ sequences per bin for a total of up to $n = 11 m$ samples per feature.  
\end{enumerate}
Observe that while the \emph{maximum} number of samples $n$ might vary across methods, in practice these bounds are never achieved due to the sparsity of SAE features. In particular, SAE features that encode interesting monosemantic features will by construction activate on very few sequences, so this is not an issue.

\textbf{SAE Activation Thresholding.}
The goal of an annotation method is to predict whether an SAE feature activates at a given residue and in order to do this we need to binarize the normalized activation across a feature labeling active/inactive. Again the method varies slightly between the two annotation method regimes:
\begin{enumerate}
    \item For continuous methods, we say that a feature $j$ fires on a residue if activation of $j$ on the residue exceeds a feature-specified threshold $\gamma$. Specifically, to define $\gamma$, we collect all non-zero residue activations at $j$ across each sampled protein and set $\gamma$ to the value determining the $0.8$ quantile. Following this, residues whose activation is greater than $\gamma$ are treated as `positively' activating. To validate predictability and train our model we sample both `background' protein residues as those who have zero activation and those whose activation is simply below the threshold (acting as hard-negatives).
    \item Unlike the continuous methods which are evaluated using PR-AUC values and hence require a fixed activation threshold per feature, the binary annotation methods are evaluated using F1 scores allowing us to sweep activation thresholds rather than fix one. For each candidate label, e.g. InterPro label, that is valid for at minimum number protein sequences, we pool the corresponding per-residue activation vectors. We then sweep a grid of threshold values consisting of $100$ equally spaced values in $[0,\text{feat max}]$ and choose the threshold value that maximizes the label's F1 value.
\end{enumerate}

\subsection{Details for Geometric Annotation} \label{sec:appendix_geometric_sampling}
\textbf{Sampling strategy.} For a given SAE feature, we sample up to $500$ of the top-activating proteins with length $n\geq21$. We split samples into three buckets as activated, hard-negatives and true-zero; for activation $x$ at a given residue, these are defined via our threshold for activation ($x > 0.8$), ($0.8> x > 0$) and $x=0$ respectively. We stress that this is an upper bound, and sparse features of intrest almost never hit this bound. 

\textbf{Thresholding strategy.} We define residues as `activated' for a feature $j$ if the corresponding SAE activation exceeds a feature-specified threshold $\gamma$. Specifically, to define $\gamma$, we collect all non-zero residue activations at $j$ across each sampled protein and set $\gamma$ to the value determining the $0.8$ quantile. Following this, residues whose activation is greater than $\gamma$ are treated as `positively' activating. To validate predictability and train our model we sample both `background' protein residues as those who have zero activation and those whose activation is simply below the threshold (acting as hard-negatives).

\subsection{Details for Residue Level Database Methods}\label{sec:appendix-residue-db}
\textbf{Sampling strategy.} For a given SAE feature, we normalize the feature's per-protein maximum activation by its global maximum across the proteome and stratify proteins into $11$ bins: one inactive bin (activation = $0$) and ten equal-width bins $[0.0, 0.1),\dots, [0.9, 1.0]$. We sample up to 50 proteins per bin for a total of up to 550 proteins per feature. 

\textbf{SAE activation thresholding.} We sweep a per-code threshold grid. For each candidate InterPro code carried by at least $3$ proteins, we pool the per-residue activation vectors of the proteins carried by the InterPro code. We then sweep a grid of threshold values consisting of $100$ equally spaced values in $[0,\text{feat max}]$ and choose the threshold value that maximizes the code's F1 value.

\subsection{Details for Sequence Motif Methods (MEME)}\label{sec:appendix-meme}
\textbf{Sampling strategy.} We sample up to $60$ proteins per SAE feature including the top $20$ highest activating proteins and up to $10$ from $4$ bins of normalized activation levels. For each of these proteins, we select $3$ highest activating residue positions and extract local windows with $h=7$ that are used as a training set. The window lengths are smaller than in the geometry method, because motif patterns longer than $15$ residues are rare. Observe that we use the full sequence lengths to compute background amino acid frequencies for the log-odds matrix and to evaluate any discovered motifs.

\textbf{Thresholding strategy.} Identical to \autoref{sec:appendix_geometric_sampling}.

\subsection{Details for Position Annotations}\label{sec:appendix-position-annotations}
\textbf{Sampling strategy.} We sample up to $60$ proteins per SAE feature including the top $20$ highest activating proteins and up to $10$ from $4$ bins of normalized activation levels.

\textbf{Thresholding strategy.} Identical to \autoref{sec:appendix-residue-db}.

\section{Binarization of Continuous Annotation Methods} \label{sec:appendix-binarization}

\textbf{GBM Geometry Score.}
To binarize the GBM predicted geometry score (specifically for evaluation via the visualizer and binarized metrics); we choose a threshold determined via 5-fold CV splits and choosing the threshold that returned highest F1 score. This helps us determine the optimal threshold within the GBM for accessing metrics such as TP/FP/TN/FN across a given sequence.

\textbf{MEME Sequence Method.}
For the MEME annotation, residue-level SAE activations are binarized by thresholding at the 80th percentile of activations across valid positions (those with finite PWM log-odds scores); if the feature is sufficiently sparse that this percentile collapses to zero, all nonzero-activation residues are instead labelled positive. The continuous PWM log-odds score at each residue is then evaluated against these binary labels using PR-AUC, matching the threshold-free evaluation procedure used for the geometry annotation.

\clearpage
\section{Geometric annotation methods}\label{sec:appendix-geometric-feats}

\subsection{GBM hyperparameters}

\begin{table}[!h]
\centering
\label{tab:gbc_hyperparameters}
\begin{tabular}{ll}
\hline
\textbf{Hyperparameter} & \textbf{Value} \\
\hline
Number of estimators & 80 \\
Maximum tree depth & 3 \\
Learning rate & 0.1 \\
Subsample ratio & 0.8 \\
Minimum samples per leaf & $\max(5, \lfloor 0.02 \times N \rfloor)$ \\
\hline
\end{tabular}
\caption{Hyperparameter configuration for the Gradient Boosting Classifier.}
\end{table}

\subsection{Local features}
The geometric features used to train the GBMs to infer geometric motifs occurring at given SAE features are defined as below; recall, we define these features across two chosen windows for each residue $i$ such that $W_{h}(i) = \{i-h, ..., i+h \}$ for $h=10$. Here, the geometric features are defined along the $\text{C}_{\alpha}$-backbone of a protein, which is defined computationally as a set of edges and vertices in $\mathbb{R}^{3}$. Specifically, we define $c_{i}\in \mathbb{R}^{3}$ to indicate the $i^{\text{th}}$ vertex, with edges connecting vertices $c_{i-1}$ and $c_{i+1}$.

\subsubsection{Curvature}
To define curvature along the $\text{C}_{\alpha}$-backbone we first determine a tangent vector at a vertex $x_{i}$,
$$T_{i} = \frac{c_{i+1}-c_{i-1}}{||c_{i+1}-c_{i-1}||}$$
From which we can define the curvature $\kappa_{i}$ as
$$\kappa_{i} = ||T_{i-1}\times T_{i}||$$

\subsubsection{Torsion}
To assign torsion, or `signed dihedral torsion' to residue $i$ we consider 4 points $(c_{i-2}, c_{i-1}, c_{i}, c_{i+1})$ from which we define 
$$b_{0} = c_{i-1} - c_{i-2}, \hspace{1em} b_{1} = c_{i} - c_{i-1}, \hspace{1em} b_{2} = c_{i+1} - c_{i}$$
Describing the 3 bond vectors between the 4 chosen points. The central bond vector is normalized, $\hat{b}_{1}$, and the adjacent bond vectors projected onto the plane perpendicular to $\hat{b}_{1}$
$$v = b_{0} - (b_{0}\cdot\hat{b}_{1})\hat{b}_{1}, \hspace{1em} w = b_{2} - (b_{2}\cdot\hat{b}_{1})\hat{b}_{1}$$
$v$ and $w$ can be seen as incoming and outgoing segments, perpendicular to $b_{1}$, the dihedral is then defined as the signed angle from $v$ to $w$ around the axis $b_{1}$.
$$x = (v\cdot w), \hspace{1em} y = (\hat{b}_{1}\times v)\cdot w$$
And so we define signed dihedral torsion at residue $i$ as 
$$\tau_{i} = \text{atan}2(y, x)$$

\subsubsection{Planarity}
The local planarity of a residue is defined by considering the covariance matrix $C_{i}$ of residues between a given window $W_{h}(i)$. From this, the fraction between minimal eigenvalue and the trace of $C_{i}$ defines a planarity score at $i$.
$$\pi_{i} = \frac{\lambda_{\text{min}}(C_{i})}{\text{tr}(C_{i})}$$

\subsubsection{Geometry statistics at different scales}

With primitive geometric features as defined above, we construct a list of feature statistics at a range of scales for maximal context which we define below.

For a given window $W_{p}$, we define the mean curvature $\mu(\kappa_{i})$, maximum curvature $\text{max}(\kappa_{i})$ and curvature standard deviation $\sigma(\kappa_{i})$. Similarly, torsion mean $\mu(\tau_{i})$, torsion standard deviation  $\sigma(\tau_{i})$, planarity mean $\mu(\pi_{i})$, and  planarity standard deviation $\sigma(\pi_{i})$.

In addition, we consider finer splits across the window $W_{p}$ considering feature windows split into N, center, and C thirds, each assigned a mean curvature, planarity, and torsion. And coarser fits with feature splits at $h/2$ and $2h$ corresponding to narrow and wide windows, each assigned mean curvature, planarity, and torsion.

We can also define the `end to end ratio' as
$$E = \frac{||c_{p+h} - c_{p-h}||}{\sum_{i=p-h}^{p+h-1} ||c_{i+1}-c_{i}||}$$

And the tangential alignment of the proteins, which measures the backbones consistency to stay along a straight path, 
$$\nu_{i} = \frac{1}{|W-1|}\sum_{i\in W}T_{i}\cdot T_{i+1}$$

This range of features across window size helps mitigate the fact that choosing a relevant window at which geometric features describe a motif is non-trivial and may vary from motif to motif.

\subsubsection{Contact statistics}\label{sec:appendix-contacts}

When considering contact residues with respect to a chosen window $W_{p}$ we exclude the residues $i \in W_{p}$. Specifically for outside residues $j$, let $d_{j} = ||c_{j} - c_{p}||$ and $s_{j} = j-p $.
From the definitions above we can consider the contact density at some distance $X$\r{A} as
$$\rho_{X} = \sum_{j\notin W_{p}} 1, \hspace{1em} \text{for } [d_{j}<X]$$
We can also extend this definition to compute the number of long range contacts at some distance $R$ away,
$$\ell_{X} = \sum_{j\notin W_{p}} 1, \hspace{1em} \text{for } [d_{j}<X \text{ and } s_{j}>R]$$

Specifically we restrict our features across $X=8, 12$ and $R=12$.

\subsubsection{Amino-acid covariates}
We have 5 measures of the amino acid composition of a given residue, computed explicitly as a fraction of amino acids with similar structural behavior.

Specifically, for a given residue $i$ with window $W_{h}(i)$ the fraction:

denote the valid residue indices in the local window, and let $s_q$ denote the amino
acid at position $q$. For any amino-acid set $A\subseteq\mathcal{A}$, define

$$
\operatorname{frac}_{A}(i)
=
\frac{1}{|W_h(i)|}
\sum_{q\in W_h(i)}
1, \hspace{1em} \text{for }\{s_q\in A\}.
$$

We use the following amino-acid groups, capturing protein flexibility, packing density, and distant structural interactions:
\[
A_{\mathrm{GP}}=\{G,P\},
\]
\[
A_{\mathrm{hydrophobic}}=\{A,V,I,L,M,F,W,Y\},
\]
\[
A_{\mathrm{charged}}=\{D,E,K,R,H\},
\]
\[
A_{\mathrm{polar}}=\{S,T,N,Q,C\},
\]
\[
A_{\mathrm{aromatic}}=\{F,W,Y,H\}.
\]

\section{Protein level baselines}\label{sec:appendix-plbaseline}

Our first two methods are very similar to \cite{silberg_towards_2025} and work on a protein level. On a given SAE feature, the InterPro annotation method at this level asks the following question: do the proteins where this SAE feature fires coincide with proteins that carry a known biological/structural annotation from one of the component databases in InterPro? Concretely, we normalize the feature's per-protein maximum activation by its global maximum across the proteome and stratify proteins into $11$ bins: one inactive bin (activation = $0$) and ten equal-width bins $[0.0, 0.1), …, [0.9, 1.0]$. We sample up to 50 proteins per bin for a total of up to 550 proteins per feature. For every InterPro code carried by at least ten of these proteins, we test whether presence of that code predicts whether the feature fires on the protein, where a protein is called "active" if the feature's activation at any of its residues exceeds a cutoff $\tau$. In order to avoid arbitrarily setting a cutoff for "high" activation, we simply sweep over a grid of candidate thresholds and report the best F1 score achieved. Per SAE feature, we determine a grid of possible thresholds and score each InterPro code on predicting whether a protein falls on either end of the threshold. We choose the code that achieves the highest F1 score across the different choices of threshold and report both the F1 value and the chosen threshold.

\textbf{Evaluation of Protein-Level Methods.} For the protein-level annotation methods, i.e. InterPro protein level, the only portion of the procedure that changes is what we permute. These methods only attempt to predict whether a SAE feature activates on a protein and not where along the protein it activates. Concretely, the input to these annotation methods is a vector of the max activation on the sampled proteins. This is the vector we shuffle to capture the null hypotesis that the proteins the SAE feature activates on carries no information of the annotation of the proteins. We then compute raw $p$ values for each annotation method on each feature and then apply the BH correction each annotation method independentely analagously to the residue level methods and label annotation method significant for a given SAE feature if its $q$ value is below $0.05$. 

\clearpage
\section{Position Predicates} \label{sec:appendix-position-predicates}
\begin{table}[h]
\centering
\small
\renewcommand{\arraystretch}{1.15}
\begin{tabularx}{\textwidth}{l l X}
\toprule
\textbf{Predicate name} & \textbf{Position condition} & \textbf{Description} \\
\midrule
\texttt{first\_5}        & \texttt{pos < 5}                          & First 5 residues (0--4). \\
\texttt{first\_10}       & \texttt{pos < 10}                         & First 10 residues (0--9).\\
\texttt{first\_20}       & \texttt{pos < 20}                         & First 20 residues (0--19); typical signal-peptide window. \\
\texttt{last\_5}         & \texttt{pos >= slen - 5}                  & Last 5 residues. \\
\texttt{last\_10}        & \texttt{pos >= slen - 10}                 & Last 10 residues. \\
\texttt{last\_20}        & \texttt{pos >= slen - 20}                 & Last 20 residues.\\
\texttt{pct\_0\_10}      & \texttt{pos / slen < 0.1}                 & First decile of the sequence (length-normalised). \\
\texttt{pct\_10\_20}     & \texttt{0.1 <= pos / slen < 0.2}          & Second decile. \\
\texttt{pct\_20\_30}     & \texttt{0.2 <= pos / slen < 0.3}          & Third decile. \\
\texttt{pct\_30\_40}     & \texttt{0.3 <= pos / slen < 0.4}          & Fourth decile. \\
\texttt{pct\_40\_50}     & \texttt{0.4 <= pos / slen < 0.5}          & Fifth decile. \\
\texttt{pct\_50\_60}     & \texttt{0.5 <= pos / slen < 0.6}          & Sixth decile. \\
\texttt{pct\_60\_70}     & \texttt{0.6 <= pos / slen < 0.7}          & Seventh decile. \\
\texttt{pct\_70\_80}     & \texttt{0.7 <= pos / slen < 0.8}          & Eighth decile. \\
\texttt{pct\_80\_90}     & \texttt{0.8 <= pos / slen < 0.9}          & Ninth decile. \\
\texttt{pct\_90\_100}    & \texttt{pos / slen >= 0.9}                & Final decile. \\
\texttt{third\_N}        & \texttt{pos / slen < 1/3}                 & N-terminal third of the sequence. \\
\texttt{third\_M}        & \texttt{1/3 <= pos / slen < 2/3}          & Middle third of the sequence. \\
\texttt{third\_C}        & \texttt{pos / slen >= 2/3}                & C-terminal third of the sequence. \\
\texttt{terminal\_10pct} & \texttt{pos / slen < 0.1 or pos / slen >= 0.9} & Either tail (first-or-last decile). \\
\texttt{interior\_80pct} & \texttt{0.1 <= pos / slen < 0.9}          & Sequence interior (complement of \texttt{terminal\_10pct}). \\
\texttt{mid\_20pct}      & \texttt{0.4 <= pos / slen < 0.6}          & Central 20\% window around the midpoint. \\
\bottomrule
\end{tabularx}
\caption{Details of the various position predicates that capture whether an SAE feature activates on a a certain position along the protein sequence. Note that $\mathtt{pos}$ is the $0$ indexed residue position, while $\mathtt{slen}$ is the length of the protein sequence, so the conditions describe the positions that a particular predicate evaluates to true.}
\label{tab:position-predicates}
\end{table}

\clearpage
\section{Geometric Feature Importance Breakdown}\label{sec:appendix-geom-feat-importance}

\begin{figure}[!h]
    \centering
    \includegraphics[width=0.93\linewidth]{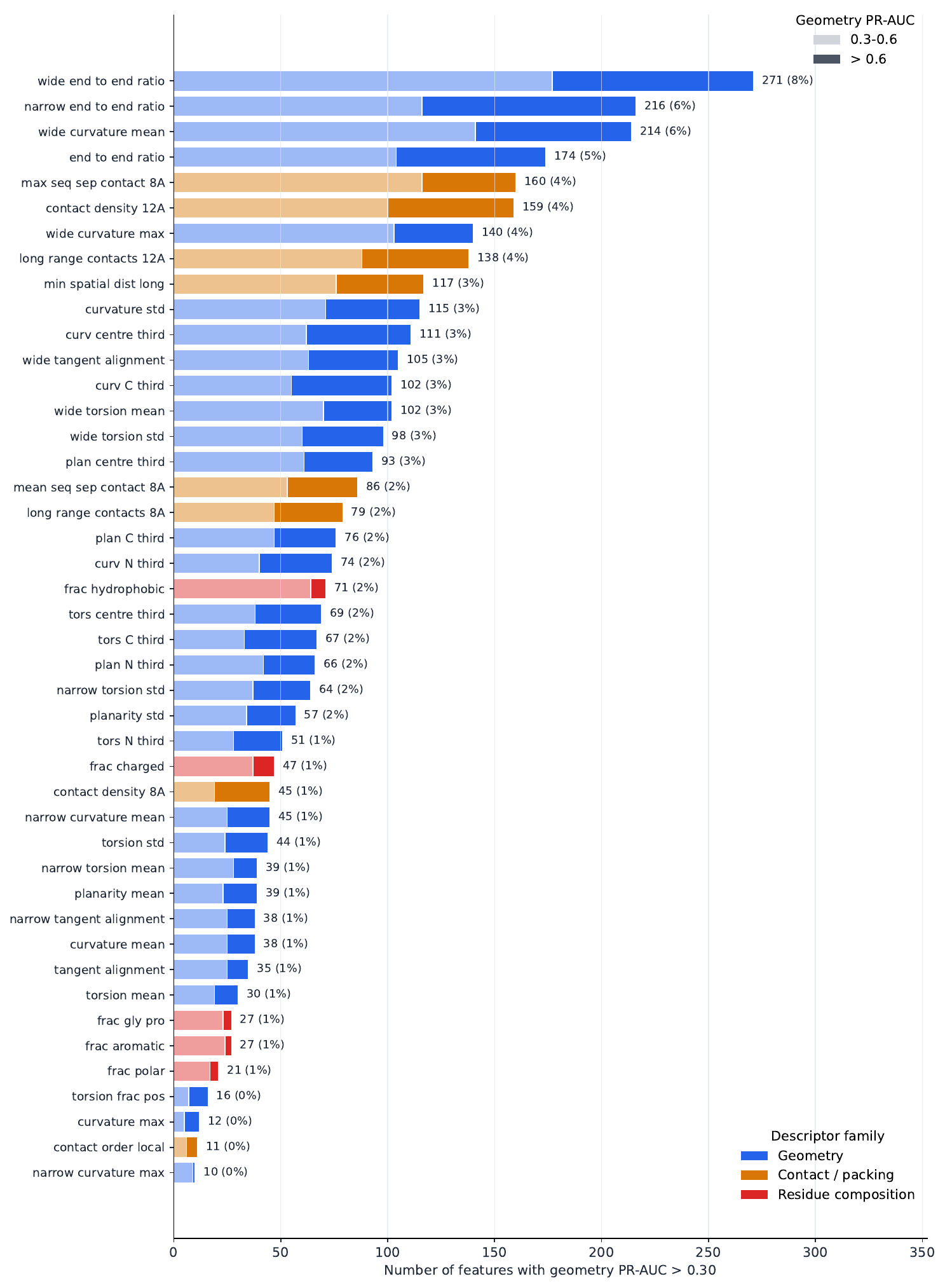}
    \caption{\textbf{Top descriptor counts at layer 4.} Top descriptor counts (GBM importance above 0.1) counted over all geometric annotations with PR-AUC$>0.3$ showing vast majority of geometric annotations exhibit minimal dependance on amino acid covariates.}
    \label{fig:appendix-breakdown}
\end{figure}

\clearpage
\section{Extension of runs to ESM-2 35M, layer 6}\label{sec:appendix-35M}

To provide an initial assessment of generalization beyond ESM-2 8M, we repeated the analysis using layer 6 of ESM-2 35M. Layer 6 is the middle layer of ESM-2 35M (which has 12 layers), which allows for a natural comparison with the layer 4 analysis of ESM-2 8M (which has 8 layers). The resulting annotation coverage and PR-AUC distribution are broadly consistent with those observed for the 8M model.

    \begin{table}[hbt!]
    \centering
    \resizebox{\textwidth}{!}{%
    \begin{tabular}{c|c|c|c|c|c|c}
         & \% Total annotated & \% InterPro Prot. & \% InterPro Res. &\% Seq Pos & \% Seq Motif & \% Geometric \\
         Layer 6 & 92.99 & 65.68 & 64.28 & 81.52 & 70.08 &  92.92 \\
         
    \end{tabular}
    }
    \caption{ FDR-controlled discovery coverage across annotation methods.
    We say that a feature is annotated by a given annotation method, if its $q$ value if below $0.05$. We say that a feature is annotated if any of the annotation methods have a $q$ value if below $0.05$.}
    \label{tab:annotated_percentage-35M}
\end{table}

    \begin{table}[hbt!]
        \centering
        \begin{tabular}{c|c|c|c}
            & PR-AUC 0.0-0.3 & PR-AUC 0.3-0.6 &  PR-AUC $>0.6$ \\
             Layer 6 & 63.20 & 26.30 & 10.50 \\
        \end{tabular}
        \caption{PR-AUC values of geometrically annotated features across layer 6 within ESM-2 35M.}
        \label{tab:PR-AUC-35M}
    \end{table}

\section{Validation with Experimentally Determined Protein Structures} \label{sec:appendix-real_protein}

To assess potential circularity associated with using AlphaFold structures, we repeated the analysis for seven proteins with experimentally determined structures; PR-AUC values from the two structure sources were closely matched, suggesting that the results are not driven by the use of AlphaFold models.

\begin{table}[hbt!]
\centering
\begin{tabular}{c|c|c|c|c|c|c}
    Protein & PDB & RMSD (Å) & Feature ID & nPos & PR-AUC (AF) & PR-AUC (crystal) \\ 
     P75804 & 2g8s & 0.13 & 1498 & 182 & 0.974 & 0.965 \\
     Q46927 & 4d79 & 0.43 & 6894 & 17 & 0.703 & 0.699 \\ 
     P0AC98 & 5zug & 0.30 & 7257 & 16 & 0.725 & 0.694 \\ 
     Q47208 & 3dmy & 0.44 & 2231 & 50 & 0.492 & 0.500 \\
     P0ADR8 & 9hs1 & 1.51 & 2231 & 28 & 0.488 & 0.516 \\
     P0ABE5 & 5oc0 & 0.44 & 6603 & 39 & 0.987 & 0.960 \\ 
     Q46901 & 5h9f & 4.96 & 4441 & 12 & 0.302 & 0.283 
\end{tabular}
\caption{Comparison of PR-AUC values from layer 4 of ESM2-8M using AlphaFold (AF) and experimentally determined crystal structures. RMSD quantifies the structural difference between the AlphaFold model and corresponding PDB structure. The nPos column refers to the number of active residues for the protein-feature pair. }
\label{tab:real_protein}
\end{table}

\end{document}